\documentclass[
  journal=largetwo,
  manuscript=article-type,
  year=2024,
  volume=37,
]{cup-journal}

\usepackage{amsmath}
\usepackage[nopatch]{microtype}
\usepackage{booktabs}
\usepackage{hyperref}
\hypersetup{
    colorlinks=true,
    linkcolor=blue,
    citecolor=blue,
    filecolor=magenta,      
    urlcolor=blue,
    pdftitle={Overleaf Example},
    pdfpagemode=FullScreen,
    }
\usepackage{txfonts}
\usepackage{graphicx}
\usepackage{amssymb}
\usepackage{makecell}

\def\aap{Astron.~Astrophys.}            
\def\apjl{Astrophys.~J.}                
\def\apjs{Astrophys.~J.~Suppl.~Ser.}    

\title{A Far-Infrared Search for Planet Nine Using AKARI All-Sky Survey}

\author{Amos Y.-A. Chen}
\affiliation{Department of Physics, National Tsing Hua University, 101, Section 2. Kuang-Fu Road, Hsinchu, 30013, Taiwan}
\email[Amos Y.-A. Chen]{yuanchen@gapp.nthu.edu.tw}

\author{Tomotsugu Goto}
\affiliation{Department of Physics, National Tsing Hua University, 101, Section 2. Kuang-Fu Road, Hsinchu, 30013, Taiwan}\alsoaffiliation{Institute of Astronomy, National Tsing Hua University, 101, Section 2. Kuang-Fu Road, Hsinchu, 30013, Taiwan}

\author{Issei Yamamura}
\affiliation{Institute of Space and Astronautical Science, Japan Aerospace Exploration Agency, 3-1-1 Yoshinodai, Chuo-ku, Sagamihara, Kanagawa 252-5210, Japan}

\author{Takao Nakagawa}
\affiliation{Institute of Space and Astronautical Science, Japan Aerospace Exploration Agency, 3-1-1 Yoshinodai, Chuo-ku, Sagamihara, Kanagawa 252-5210, Japan}\alsoaffiliation{Advanced Research Laboratories, Tokyo City University, Setagaya-ku, Tokyo, JP
}

\author{Cossas K.-W. Wu}
\affiliation{Institute of Astronomy, National Tsing Hua University, 101, Section 2. Kuang-Fu Road, Hsinchu, 30013, Taiwan}

\author{Terry Long Phan}
\affiliation{Institute of Astronomy, National Tsing Hua University, 101, Section 2. Kuang-Fu Road, Hsinchu, 30013, Taiwan}

\author{Tetsuya Hashimoto}
\affiliation{Department of Physics, National Chung Hsing University, 145, Xingda Road, Taichung, 40227, Taiwan}

\author{Yuri Uno}
\affiliation{Department of Physics, National Chung Hsing University, 145, Xingda Road, Taichung, 40227, Taiwan} 

\author{Simon C.-C. Ho}
\affiliation{Research School of Astronomy and Astrophysics, The Australian National University, Canberra, ACT 2611, Australia}\alsoaffiliation{Centre for Astrophysics and Supercomputing, Swinburne University of Technology, P.O. Box 218, Hawthorn, VIC 3122, Australia}\alsoaffiliation{OzGrav: The Australian Research Council Centre of Excellence for Gravitational Wave Discovery, Hawthorn, VIC 3122, Australia}\alsoaffiliation{ASTRO3D: ARC Centre of Excellence for All-sky Astrophysics in 3D, ACT 2611, Australia}

\author{Seong Jin Kim}
\affiliation{Institute of Astronomy, National Tsing Hua University, 101, Section 2. Kuang-Fu Road, Hsinchu, 30013, Taiwan}


\begin{document}

\begin{abstract}

An unusual orbital element clustering of Kuiper belt objects (KBOs) has been observed. The most promising dynamic solution is the presence of a giant planet in the outer Solar system, Planet Nine. However, due to its extreme distance, intensive searches in optical have not been successful. We aim to find Planet Nine in the far-infrared, where it has the peak of the black body radiation, using the most sensitive all-sky far-infrared survey to date, {\it AKARI}. In contrast to optical searches, where the energy of reflected sunlight decreases by $d^{4}$, thermal radiation in the infrared decreases with the square of the heliocentric distance $d^{2}$.  
We search for moving objects in the AKARI Single Scan Detection List. We select sources from a promising region suggested by an N-body simulation from Millholland and Laughlin 2017: $30^{\circ}<$ R.A. $<50^{\circ}$ and $-20^{\circ}<$ Dec. $<20^{\circ}$. Known sources are excluded by cross-matching {\it AKARI} sources with 9 optical and infrared catalogues. Furthermore, we select sources with small background strength to avoid sources in the cirrus.
Since Planet Nine is stationary in a timescale of hours but moves on a monthly scale, our primary strategy is to select slowly moving objects that are stationary in 24 hours but not in six months, using multiple single scans by AKARI. The selected slowly moving {\it AKARI} sources are scrutinised for potential contamination from cosmic rays.
Our analysis reveals two possible Planet Nine candidates whose positions and flux are within the theoretical prediction ranges. These candidates warrant further investigation through follow-up observations to confirm the existence and properties of Planet Nine.
\end{abstract}

\section{Introduction}

The unexpected clustering of a set of distant Kuiper Belt Objects (KBOs) claimed by \cite{Batygin_2016} suggests the presence of a massive object in the outer Solar system. This object could be an unknown planet --- called Planet Nine --- or even a primordial black hole \citep{pbh}.
Numerous studies support the hypothesis of Planet Nine to account for this phenomenon.
\cite{Batygin_2016} employed an n-body simulation to model the impact of a perturber on the orbits of test particles.
They showed that the observed KBOs with such clustering of the argument of perihelion are only 0.007\% likely to occur by chance, and their simulations suggested that the existence of Planet Nine could produce this clustering.

However, there are several studies arguing that the clustering of extreme trans-Neptune Objects (TNOs) arises from the observational bias. \cite{Shankman2017} reported a uniform distribution of eight extreme TNOs discovered by the Outer Solar System Origins Survey (OSSOS; \cite{Bannister2016}) survey. \cite{Bernardinelli2020} found no significant clustering of seven extreme TNOs from the Dark Energy Survey (DES; \cite{DES}). \cite{Napier2021} examined 14 extreme TNOs reported in the OSSOS, DES, and \cite{SheppardTrujillo2016} studies, concluding that the mean scaled longitude of perihelion and orbital poles of the identified extreme TNOs align with a uniformly distributed population at a range between 17\% and 94\%. Although \cite{Napier2021} showed no clustering of TNOs, their work did not rule out the Planet Nine hypothesis due to the limited observation area.

In light of these competing interpretations, a recent paper \cite{Batygin_2024} presented a new simulation on the orbits of long-period TNOs. They correct for observational biases and compare their N-body simulation result with the perihelion distribution of 17 long-period TNOs from the Minor Planet Center database. Their result showed that the perihelion distribution of these TNOs rejects the scenario without Planet Nine at the $\sim5$-$\sigma$ confidence level. 
Several simulation papers have also provided predictions regarding the characteristics of Planet Nine.
\cite{Batygin_2016} estimated that Planet Nine is on a 700 au semimajor axis and 0.6 eccentricity orbit with 10 Earth masses ($M_{\oplus}$). Their updated simulation with a more complete setup \citep{brown2021orbit} suggested Planet Nine is on an orbit with semimajor axis $380^{+140}_{-80}$ au and perihelion distance $300^{+85}_{-60}$ au with $6.2^{+2.2}_{-1.3}$ $M_{\oplus}$. Another simulation work \citep[][hereafter ML17]{Millholland_2017} focused on the mean motion resonances between known KBOs and Planet Nine. They suggested that Planet Nine is on an orbit with a semimajor axis of 654 au and eccentricity of 0.45, separated 400 to 900 au from Earth and 6 to 12 $M_{\oplus}$.

There were various works that searched for Planet Nine. Some previous studies expected to detect reflected Sunlight from Planet Nine in optical wide-field surveys, e.g., Zwicky Transient Facility (ZTF) \citep{Brown_2022_ztfsearch}, Dark Energy Survey (DES) \citep{Belyakov_2022_dessearch}, and Pan-STARRS1 \citep{brown2024ps1search}. Other searches utilised near-infrared all-sky surveys such as the Wide-field Infrared Survey Explorer (WISE) \citep{wisesearch} and NEOWISE \citep{neowisesearch}, but those searches were not successful. \cite{Naess2021} analysed millimetre-wave data from the Atacama Cosmology Telescope (ACT) and provided 10 possible Planet Nine candidates.

In this work, we try to find the thermal radiation of Planet Nine from the {\it AKARI} far-infrared all-sky survey data. The benefit of searching in far-infrared wavelengths is that the intensity of radiation does not drop as quickly as reflected light. Reflected Sunlight will decrease by $d^4$ as the distance $d$ increases, while thermal radiation falls by $d^2$. \cite{Cowan_2016_cmb} suggested that current and planned microwave detectors would detect Planet Nine. Other pioneering works searched for thermal radiation from Planet Nine in far-infrared all-sky surveys \citep[e.g,][]{IRASsearch, C.Sedgwick&S.Serjeant_akari_irassearch}. \cite{IRASsearch} searched in InfraRed Astronomical Satellite (IRAS) data and found one candidate with estimated distance 225$\pm$15 au and 3-5 $M_{\oplus}$, which is not consistent with most of the predictions. \cite{C.Sedgwick&S.Serjeant_akari_irassearch} searched for moving objects that have detections in the IRAS catalogues and AKARI-FIS Bright Source Catalogue (FISBSC). They found 535 potential candidates after spectral energy distribution (SED) fits. Since all of them are located in the cirrus cloud, there is no good Planet Nine candidate.

\begin{figure}
    \centering
    \includegraphics[width=\columnwidth]{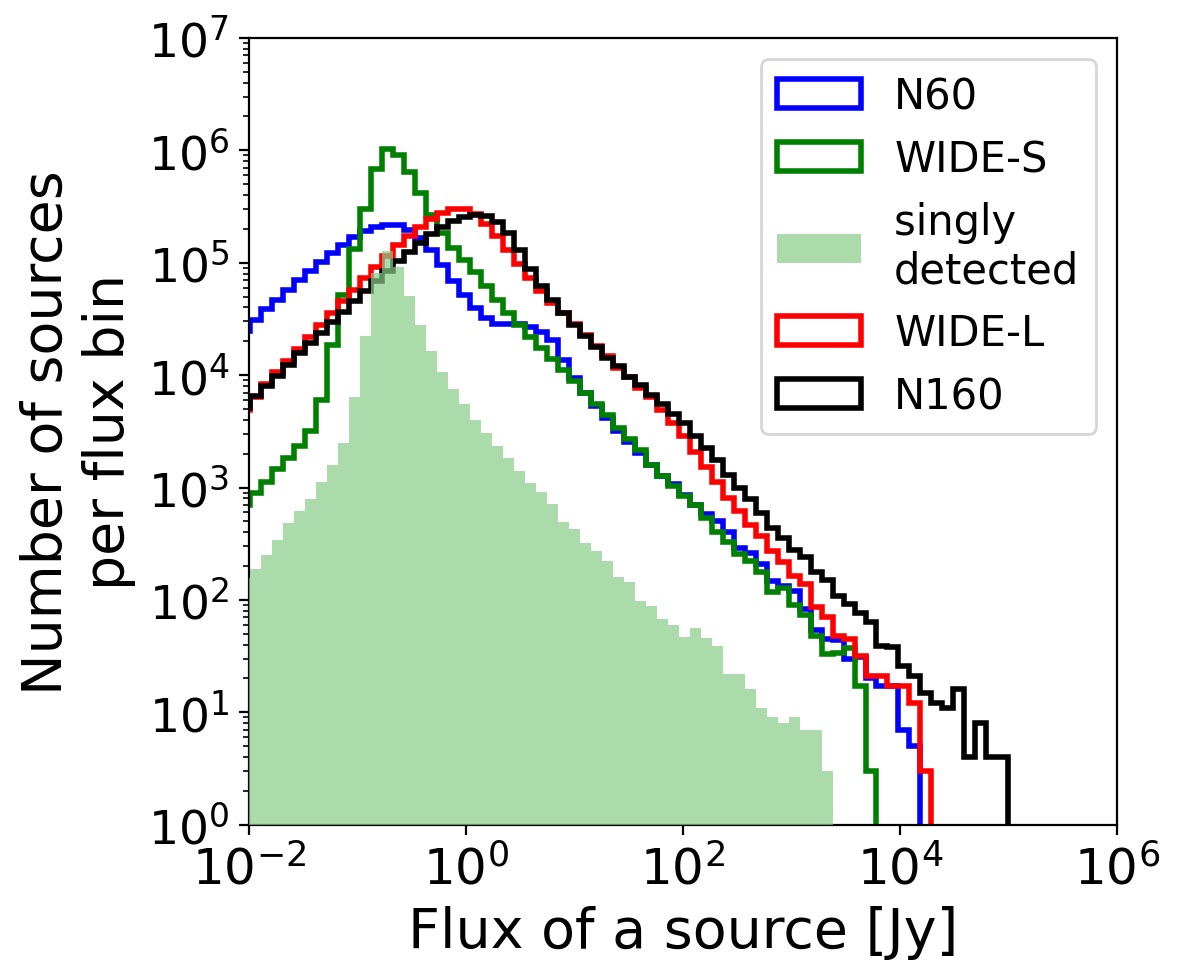}
    \caption{Flux histograms of FISSSDL sources. Four histograms correspond to four AKARI/FIS filters. The bin size is on a logarithmic scale, totalling 80 bins.
    Sources detected in the N60 filter (centred at 65\,$\mu$m) are represented in blue, those detected by the WIDE-S filter (90\,$\mu$m) are in green, the WIDE-L filter (140\,$\mu$m) results are shown in red, and the N160 filter (160\,$\mu$m) sources are depicted in black. The green shaded histogram shows the sources that were only detected once by the AKARI/WIDE-S}
    \label{fig:SDLflux}
\end{figure}

Planet Nine was estimated to have a temperature range between 53 and 28 K \citep{Cowan_2016_cmb}, corresponding to black body radiation peaks at 54.7 and 103.5 $\mu$m. {\it AKARI} Far Infrared Surveyor (FIS) instrument was equipped with four filters spanning 50 to 180\,$\mu$m \citep{Kawada2007}, which is ideal for searching for thermal emission in this temperature range. 
In this study, we use a dedicated source list "AKARI-FIS Single Scan Detection List" (hereafter FISSSDL; see Section~\ref{sec:Data}). Fig.~\ref{fig:SDLflux} shows the histogram of the FISSSDL sources ($\log N$ -- $\log S$ plot), where S in Jy is the source flux, and N is the number of objects at the flux. The most sensitive band, WIDE-S, has a peak at $\sim$0.2~Jy, lower than FISBSC Version 2 ($\sim$0.44~Jy) and IRAS Faint Source Catalogue ($\sim$1.0~Jy at 100\,$\mu$m), and includes even fainter sources. Although most of them are possibly fake sources caused by cosmic rays, there is a chance of detecting Planet Nine candidates if we carefully examine the list.
\cite{IRASsearch, C.Sedgwick&S.Serjeant_akari_irassearch} required detection in IRAS, which limited the survey to $>1.0$ Jy in $100~\mu $m. Only requiring {\it AKARI} detection can expand the search to five times fainter objects.

The paper is structured as follows. We describe the data in Section \ref{sec:Data}. Selection details in Section \ref{sec:method}. Results in Section \ref{sec:results}. Discussion and conclusion in Section \ref{sec:discussion}.


\section{AKARI-FIS Single Scan Detection List}
\label{sec:Data}

The infrared astronomical satellite \textit{AKARI} \citep{Murakami2007} was equipped with a 68.5~cm aperture cooled telescope and two science instruments: Far-Infrared Surveyor (FIS; \cite{Kawada2007}) and Infrared Camera (IRC; \cite{Onaka2007}). \textit{AKARI} was launched in 2006 and carried out the All-Sky Survey in four far-infrared wavelength bands and two mid-infrared wavelength bands as well as thousands of pointed observations of particular targets or sky areas from May 2006 to August 2007. After cryogen (liquid helium) depletion, \textit{AKARI} continued observation until February 2011, only in the near-infrared wavelengths.

Infrared point source catalogues were produced from the all-sky survey data. The AKARI-FIS Bright Source Catalog (FISBSC; Yamamura et al. 2010) includes 411 (Version 1) and 501 (Version 2) thousand sources observed in four wavelengths centred at 65\,$\mu$m, 90\,$\mu$m, 140\,$\mu$m and 160\,$\mu$m. FISBSC has already been used to search for Planet Nine \citep{C.Sedgwick&S.Serjeant_akari_irassearch}.

However, the FISBSC is not optimal for our search for Planet Nine because moving objects can be removed from the catalogue.
\textit{AKARI} scanned a position of the sky in a few to several successive orbits, then revisited the position after half a year. The number of scans depends on the Ecliptic latitude and the survey operation program, ranging from zero to several hundred. In the construction of FISBSC, a condition is set to confirm whether a signal is from a real, stationary source that it is detected in at least two scans and at least 3/4 of the total number of scans observed at the position. This condition rejects moving targets and includes Planet Nine.
Therefore, we created a dedicated source list, FISSSDL, for the Planet Nine search, from the same intermediate data used for FISBSC. FISSSDL relaxes the above confirmation condition and includes any source that was detected at least once. This change of the confirmation policy allows moving objects such as Planet Nine to be included, with the risk of contamination by many fake signals, such as cosmic rays (CRs) hits and instrumental artefacts. We carried out a careful investigation to find Planet Nine candidates, as we explain in the following sections.

\section{Methods}
\label{sec:method}

To select Planet Nine from AKARI FISSSDL, it is important to make sure that Planet Nine is bright enough to be detected by {\it AKARI}. 
Fig.~\ref{fig:SDLflux} shows a histogram of all FISSSDL sources detected in 4 filters. The AKARI/WIDE-S ($90~\mu$m) filter has a lower peak value compared to AKARI/WIDE-L and AKARI/N160. AKARI/WIDE-S detects more sources than the AKARI/N60 filter. We confirm that the WIDE-S filter is the most sensitive AKARI/FIS filter. 

Therefore, we focus on the 90\,$\mu $m flux of the sources in this work. Estimation of flux and motion of Planet Nine is described in Section~\ref{sec:Flux} and Section~\ref{sec:Motion}. In Section~\ref{sec:position}, based on the simulation results (ML17), we select candidates from the most promising area; Section~\ref{sec:xmatch}: exclude known sources by cross-match with known catalogues; Section~\ref{sec:fluxselec}: exclude sources potentially contaminated by cirrus; Section~\ref{sec:detectioin}: exclude non-moving sources; Section~\ref{sec:imageinspection}: exclude contamination from CRs and select candidates with clear detection.

\subsection{Flux Estimation}
\label{sec:Flux} 

To calculate the flux from Planet Nine, we need to know the radius of Planet Nine.
Previous simulation works \citep[e.g,][]{Batygin_2016, Brown_2019_bias, brown2021orbit, Millholland_2017} predicted the mass of Planet Nine and its orbital parameters, but there is no implication of Planet Nine's average density. Here, we assume that Planet Nine has an average density of Neptune and Uranus $\rho = (\rho_{Neptune}+\rho_{Uranus})/2 = 1.454~g/cm^3$. The size of Planet Nine can then be derived from the mass and average density. The estimated effective temperature of Planet Nine ranges from 28 to 53 K, with energy mainly from internal heat \citep{Cowan_2016_cmb}. The expected flux of black body radiation from Planet Nine at 90\,$\mu$m can be calculated as a function of distance $d$ and mass $M$ of Planet Nine:

\begin{equation}
    F = SR \frac{\Omega \lambda^2}{c}
    \label{eq1}
\end{equation}

\begin{equation}
    \Omega = \frac{\pi R^2}{d^2} = \frac{\pi}{d^2}\left(\frac{3M}{4\pi\rho}\right)^{2/3}
    \label{eq2}
\end{equation}
$F$ is flux in Jansky, $SR$ is the spectral radiance in unit $[W/m^3/sr]$, $\Omega$ is the solid angle in unit $[sr]$, $c$ is the speed of light and wavelength $\lambda$ is 90\,$\mu $m. The radius of Planet Nine $R$ can be calculated from $M$ and density $\rho$. We adopt the mass range of 6 to 12 $M_{\oplus}$ predicted by ML17. We derive the black body's spectral radiance at 90\,$\mu$m for temperatures of 28 K and 53 K suggested by \cite{Cowan_2016_cmb}; then, we plug in Eq.~\ref{eq1} to calculate the expected flux from Planet Nine. We use the mass range from ML17 to maintain consistency in this work. We adopt the promising area from ML17, so we also use the mass range predicted by ML17. We use the Planet Nine parameters estimated by ML17 rather than \cite{brown2021orbit} because 78\% of the parameter space predicted by \cite{brown2021orbit} is ruled out by recent surveys \citep{Brown_2022_ztfsearch, Belyakov_2022_dessearch, brown2024ps1search}. \cite{brown2024ps1search} combined three surveys and excluded orbits with Planet Nine brighter than 21 V magnitude, which roughly corresponds to 500 au. However, the constraints on larger distances remain weak. On the other hand, ML17 predicted a Planet Nine orbit with a more significant distance ($\sim$800 au), which has not yet been fully investigated.

\begin{figure}
    \centering
    \includegraphics[width=\columnwidth]{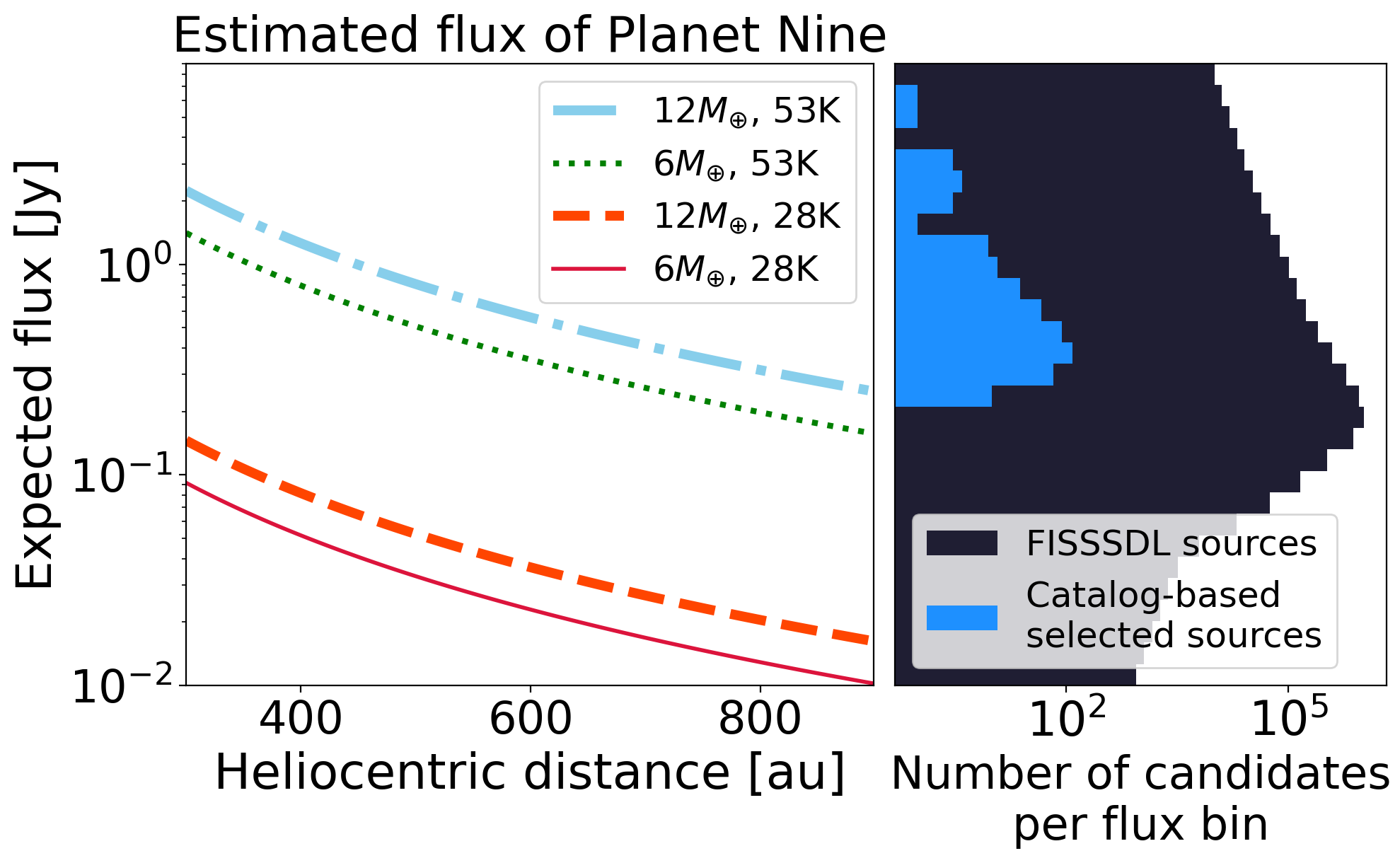}
    \caption{
    \textit{Left:} Estimated 90\,$\mu$m flux of Planet Nine. We calculate the expected Planet Nine's 90\,$\mu$m flux in Section~\ref{sec:Flux} and plot it with 4 combinations of 2 parameters: mass and temperature in the left panel. The x-axis is the heliocentric distance of Planet Nine. The mass range of 6 to 12 $M_{\oplus}$ was predicted by ML17. \cite{Cowan_2016_cmb} suggested a temperature range of 28 K to 53 K. 
    \textit{Right:} A histogram of FLUX90. The dark blue histogram shows the flux distribution of FISSSDL sources with the same y-axis as the left panel. 393 candidates selected from FISSSDL after cross-matching with known catalogues (Section~\ref{sec:xmatch}),  FLUX90/FERR90~$>3$ and BG90~$<0.2$ in catalogue unit (Section~\ref{sec:fluxselec}), and no monthly confirmation (Section~\ref{sec:fluxselec}) is shown in the light blue histogram. The X-axis shows the number of sources in each bin. The bin size is the same as Fig.~\ref{fig:SDLflux}.
    }
    \label{fig:fluxestmate}
\end{figure}

The expected flux of Planet Nine with different parameters is plotted in the left panel of Fig. \ref{fig:fluxestmate}. The right panel of Fig. \ref{fig:fluxestmate} shows the histogram of 90\,$\mu$m flux (FLUX90) of FISSSDL sources, and those sources that passed through flux selection are comparable to Planet Nine's expected flux at 53 K. \cite{brown2021orbit} used \cite{Wu2013}'s mass-radius relation derived from the Kepler planets, $M\simeq3M_{\oplus}(R/R_{\oplus})$, to estimate the radius of Planet Nine. If we adopt their mass-radius relation, the expected spectral flux drops by 50\% at 6 $M_{\oplus}$ and increases by 30\% at the 12 $M_{\oplus}$ situation, which is comparable with a constant density assumption within our interest mass range.

\subsection{Expected Motion}
\label{sec:Motion}

\begin{figure}
    \centering
    \includegraphics[width=\columnwidth]{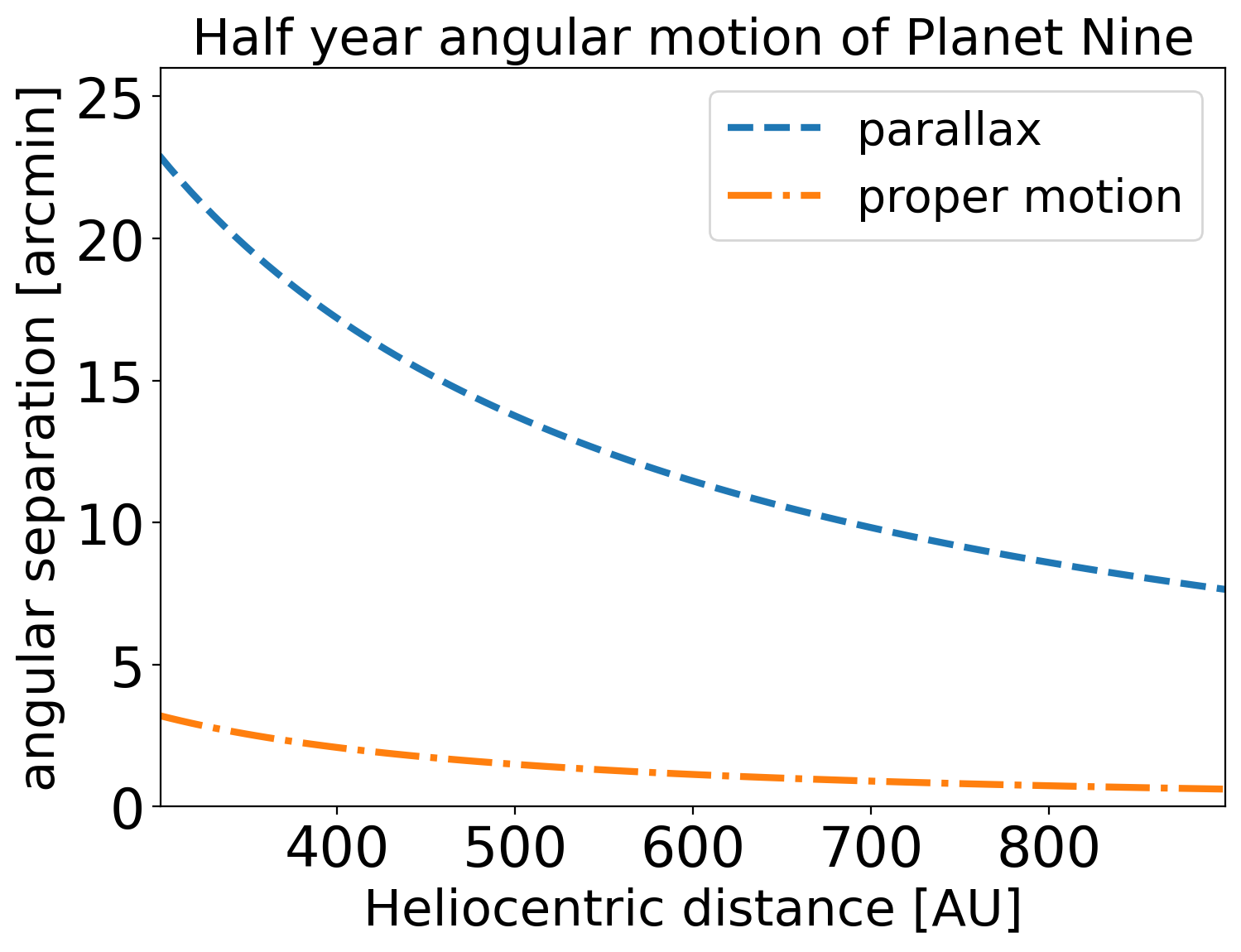}
    \caption{Anticipated proper motion and parallax of Planet Nine within half a year. The overall angular displacement is the vector sum of the proper motion and the parallax. In the time scale of half a year, parallax dominates the angular motion, so we only consider the parallax of Planet Nine in this work.}
    \label{fig:motion}
\end{figure}

The angular motion of Planet Nine is composed of proper motion and parallax, which can be calculated by equations 4 and 7 in \citep{Cowan_2016_cmb}. Their value as a function of distance to Earth is shown in Figure~\ref{fig:motion}. Planet Nine's parallax over six months ranges from 10 to 25 arcminutes, depending on its distance to the Earth. The proper motion ranges from 0.6 to 3.2 arcminutes, which is much smaller than parallax. Therefore, we only consider the parallax of Planet Nine in this work. The astrometric accuracy of the AKARI FISBSC is $3.5''$, so the parallax is detectable by {\it AKARI} with observations spaced six months apart, but not within a single day. The one-hour parallax of Planet Nine at 300 au is only $0.23''$. Therefore, we look for stationary objects in hours time scale, but moving in months time scale as good candidates for Planet Nine.

\subsection{Position Selection}
\label{sec:position}

The dynamical simulation from ML17 suggests that the probability of finding Planet Nine is higher in the region $30^{\circ} < \text{R.A.} < 50^{\circ},~-20^{\circ} < \text{Dec.} < 20^{\circ}$. This region also overlaps with the area suggested by \cite{brown2021orbit}. There are 50,033 sources in this area out of 5,274,338 sources in FISSSDL.

\subsection{Cross-match with 9 Catalogues}
\label{sec:xmatch}

Some stationary objects or known Solar-system objects can be included in the FISSSDL. To exclude known sources, we cross-match AKARI FISSSDL with 9 external catalogues: 2MASS, NOMAD, Pan-STARRS DR1(PS1), WISE, ALLWISE, CatWISE2020, unWISE, SIMBAD, and SDSS DR16.
Their properties are described in Table~\ref{tab:cats_table}. Because these observations are carried out in different years, if an object is found in multiple catalogues at the same position, it is not a moving object and thus not Planet Nine.

\begin{table*}
	\centering
         \caption{
         Catalogues used for cross-matching with AKARI FISSSDL. All data we used for cross-matching were accessed from the CDS cross-matching service.
         }
	\label{tab:cats_table}
	\begin{tabular}{cccccc}
		\hline
		Catalog Name & Number of Sources & Filters & Reference & Cross-match Radius ($''$)  \\
            \hline
		2MASS & 470,992,970 & J, H, K  & \citep{2mass} & 7.1 \\
            \hline
		NOMAD & 1,117,612,732 & B, V, R, J, H, and K & \citep{nomad} & 5.3 \\
            \hline
		PS1 & 1,919,106,885 & g, r, i, z, and y & \citep{ps1} & 4.5 \\
            \hline
		WISE & 563,921,584 & 3.4, 4.6, 12, and 22 $\mu $m & \citep{wise} & 7.9 \\
		  \hline
		ALLWISE & 747,634,026 & 3.4, 4.6, 12, and 22 $\mu $m & \citep{neowise} & 8.0 \\
		  \hline
		unWISE & 2,214,734,224 & 3.4, 4.6, 12, and 22 $\mu $m & \citep{unwise} & 4.9 \\
		  \hline
		CatWISE2020 & 1,890,715,640 & 3.4, 4.6, 12, and 22 $\mu $m & \citep{catwise} & 5.1 \\
		  \hline
		SIMBAD & 17,959,486 & B,V,R,J,H,K,u,g,r,i, and z & \citep{simbad} & 3.7 \\
		  \hline
		SDSS DR16 & 1,231,051,050 & u,g,r,i, and z & \citep{sdss} & 4.9 \\
		  \hline
		  \hline
            AKARI FISSSDL & 5,274,338 & 65, 90, 140 and 160 $\mu $m & & \\
		\hline
	\end{tabular}
\end{table*}

\begin{figure*}
    \centering
    \includegraphics[width = 0.9\columnwidth]{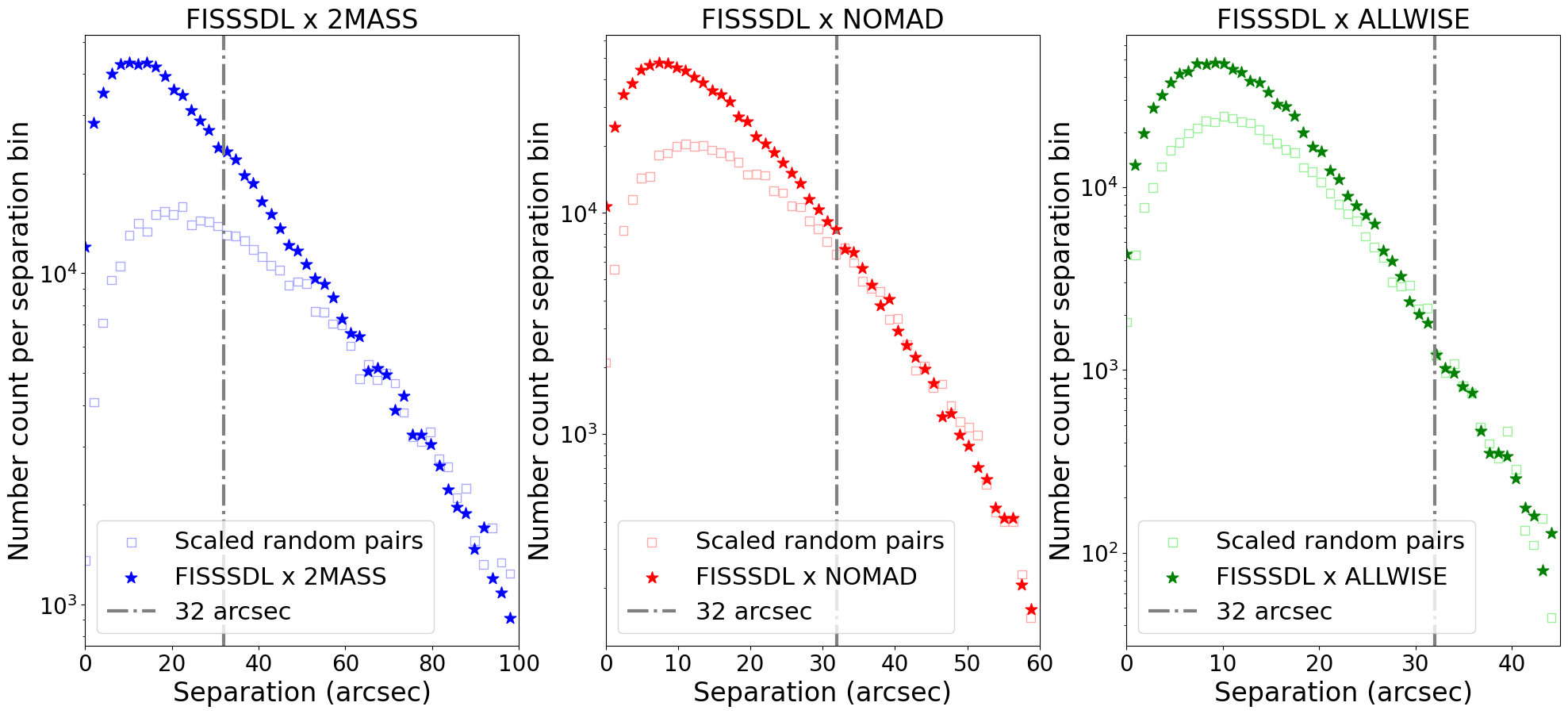}
    \includegraphics[width = 0.9\columnwidth]{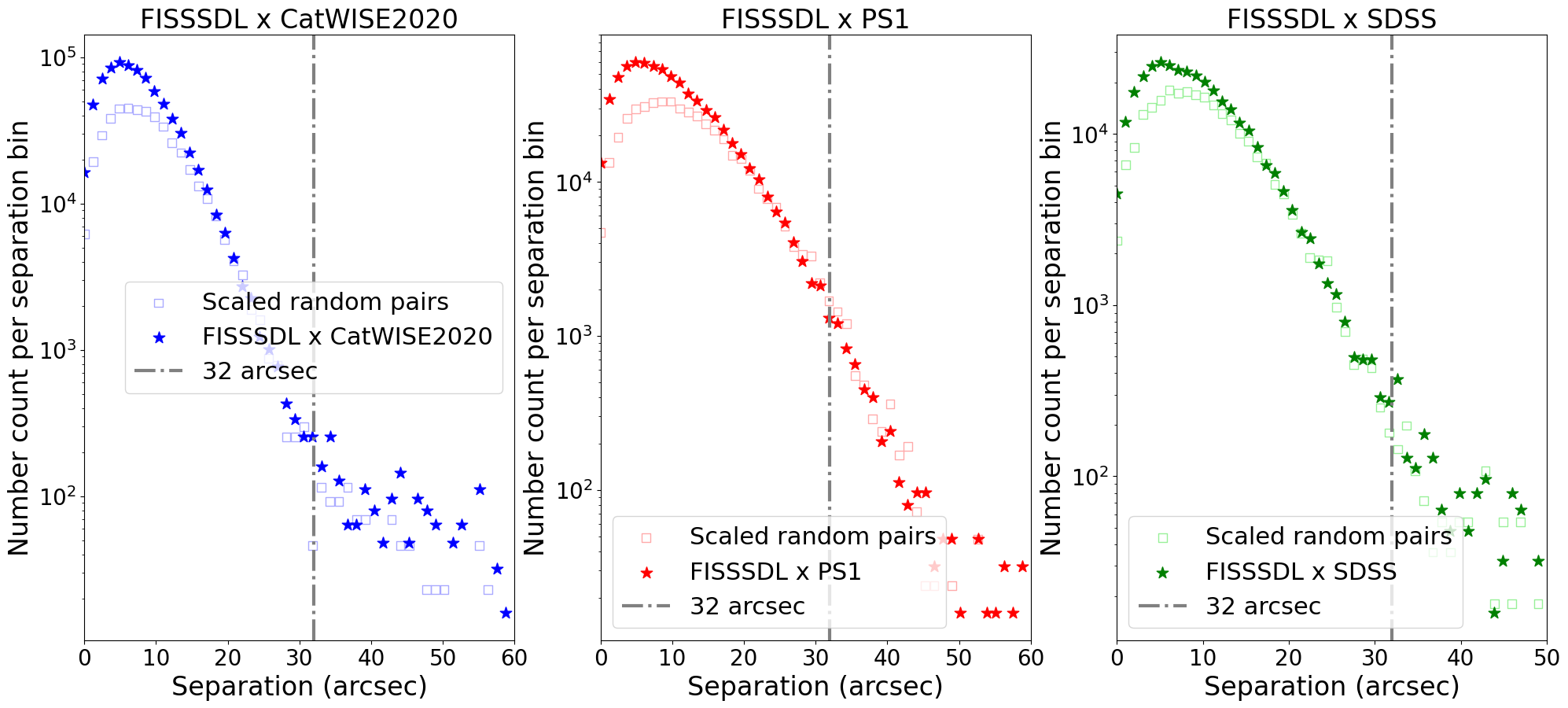}
    \includegraphics[width = 0.9\columnwidth]{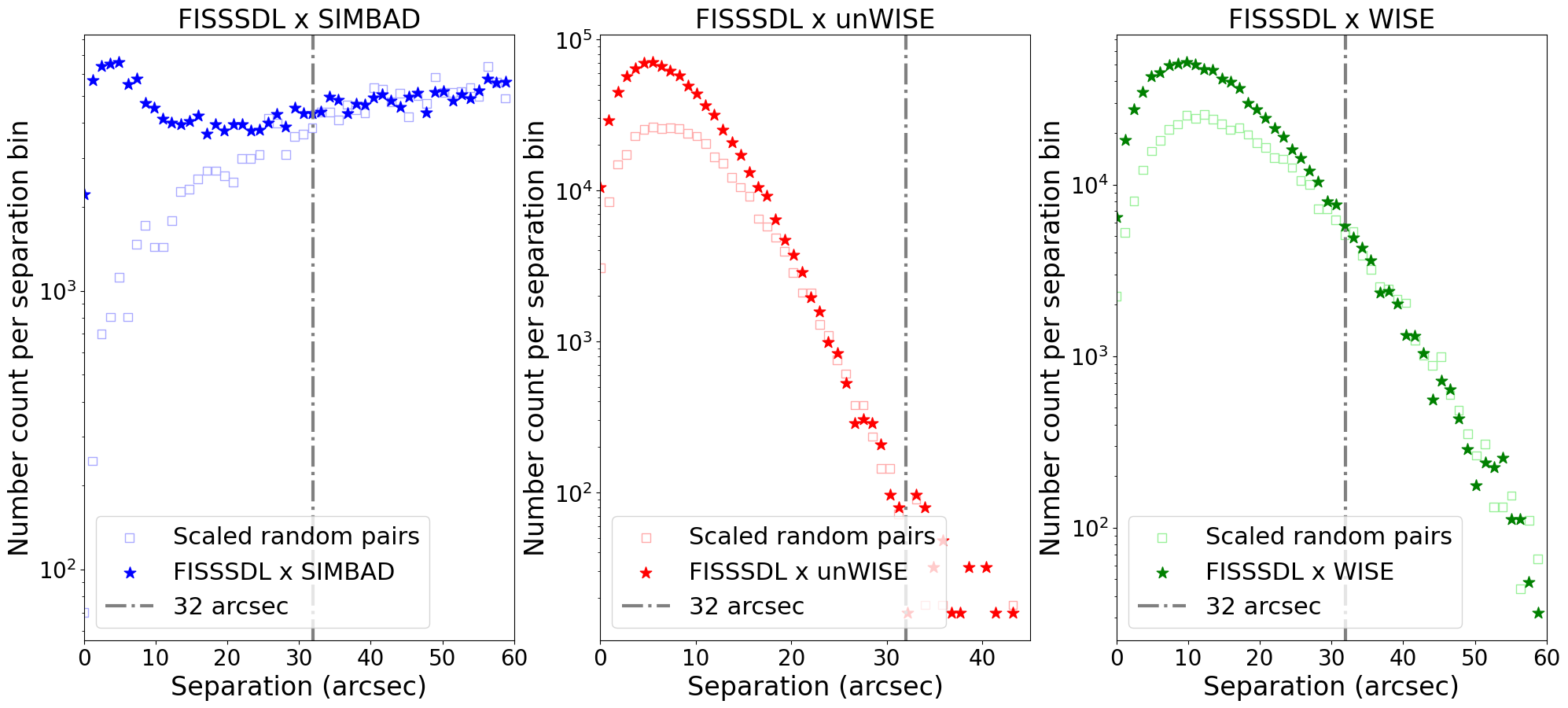}
    \caption{Separation distribution of matched sources. There are 50 bins on a linear scale in each subplot. The stars show the separation of matched pairs of {\it AKARI} sources and sources from catalogues in Table~\ref{tab:cats_table}. Squares represent the scaled separation between matched random sources and sources from the corresponding catalogue. The vertical dash line represents $32''$, beyond which two sources are treated as distinct. (see Section~\ref{sec:xmatch}).}
    \label{fig:separation}
\end{figure*}

The cross-matching of AKARI FISSSDL with the other 9 catalogues listed in Table \ref{tab:cats_table} aims to exclude stable sources. However, due to spurious sources in the FISSSDL, there are randomly matched pairs in the cross-matching result. 
To model the distribution of random sources and avoid the effect from the galactic plane, we generate 20000 random sources with a uniform distribution across the region galactic latitude ($b$) $b>10^{\circ}$ and cross-match with 9 target catalogues. We cross-match $b>10^{\circ}$ FISSSDL sources with those 9 catalogues and find the separation distribution of real and randomly matched sources. 
Sources that are at least $32''$ apart are considered distinct by definition in the point source catalogue processing (Yamamura et al. 2010). Thus, randomly matched pairs should dominate the separation distribution larger than $32''$. We fit the FISSSDL cross-match distribution at separations larger than $32''$ with the distribution of random pairs and find the portion of randomly matched sources (see Fig. \ref{fig:separation}). After subtracting the contribution from the randomly matched pairs, we are able to fit the separation distribution of the real matched pairs with the Gaussian distribution. The Gaussian fitting results are shown in the last column of Table~\ref{tab:cats_table}. The 1-$\sigma$ radius from this Gaussian fit is taken as the cross-match radius for the respective catalogue. Suppose the distance between an {\it AKARI} source and its corresponding source in a catalogue is less than the cross-match radius specific to that catalogue; we consider them to be the same object and remove this {\it AKARI} source. After this process, 29901 sources remain.

To exclude possible IRAS sources included in AKARI FISSSDL, we try to apply a similar analysis to IRAS's catalogues. However, due to the small number of sources in IRAS catalogues (245,889 sources in the IRAS Point Source Catalog (PSC) and 173,044 sources in the IRAS Faint Source Catalog (FSC), which is 3 to 4 orders less than other catalogues listed in Table~\ref{tab:cats_table}), we are not able to fit the separation distribution of matched FISSSDL IRAS sources with the random pairs at $32''<$ separation $<120''$. Thus, despite the similarity of IRAS and {\it AKARI}, we do not cross-match between FISSSDL and IRAS's catalogues. We use the positional uncertainty from IRAS PSC and IRAS FSC, and only 53 out of 29901 sources are matched. The effect of not cross-matching with IRAS is very minor, and those candidates are also excluded in the flux selection step.

\subsection{Flux Selection}
\label{sec:fluxselec}

To ensure reliable flux measurements unaffected by the cirrus effect, we utilize the index of background strength at 90\,$\mu$m (BG90, in arbitrary unit)\footnote{We use the face values in the catalogue, where the unit is not specified yet because the background values are not calibrated.} from the AKARI FISSSDL to identify candidates. In the FLUX90-BG90 plot Figure~\ref{fig:fluxbg}, the number of sources peaks at BG90~$=0.3$, and we find that most of these sources come from the cirrus region in our target area. To remove sources contaminated by cirrus, we select sources with BG90~$<0.2$ ($\sim 4$ MJy/SR). We also require flux over flux error (FERR90) larger than 5 at 90\,$\mu$m (FLUX90/FERR90~$>5$) to select sources with reliable flux. In this step, 1726 sources out of 29901 sources are selected.

\subsection{Detection Selection}
\label{sec:detectioin}
\begin{figure}
    \centering
    \includegraphics[width=\columnwidth]{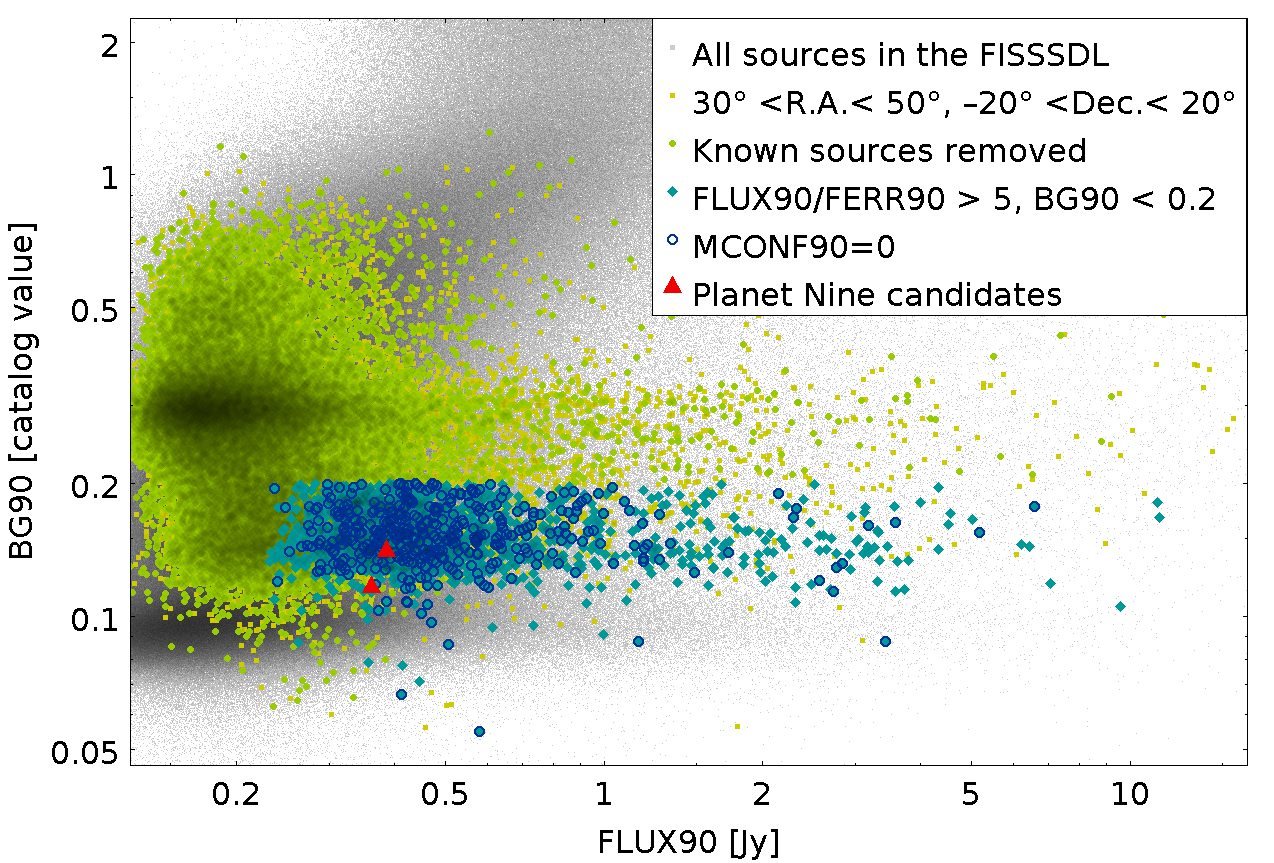}
    \caption{Candidates remained after each step on the 90\,$\mu $m flux-background face value plot. All sources from FISSSDL are marked with grey dots. Sources in the region $30^{\circ} < \text{R.A.} < 50^{\circ},~-20^{\circ} < \text{Dec.} < 20^{\circ}$ are marked with yellow squares. Green circles are sources left after removing known sources by cross-matching with 9 catalogues (See Section~\ref{sec:xmatch}). Light blue diamonds are FLUX90/FERR90~$>5$, BG90~$<0.2$ sources. The dark blue rings are sources with no monthly confirmation at 90\,$\mu$m (MCONF90=0). Two Planet Nine candidates are shown in red triangles. The FLUX90 of these candidates are catalogue fluxes, so it is different from the per scan flux.}
    \label{fig:fluxbg}
\end{figure}

Based on the discussion in Section~\ref{sec:Motion}, {\it AKARI} should be able to detect six months' parallax motion of Planet Nine if it is located from 300 to 900 au.
The FISBSC, as well as FISSSDL, contain information on whether the source is detected at periods separated by six months (Monthly confirmation, MCONF90=0 when the source does not have monthly confirmation at 90\,$\mu$m). We exclude sources confirmed over monthly intervals as they are not moving sources. We also remove sources only detected once in the 90\,$\mu$m (WIDE-S) band. 393 candidates passed these selection criteria.
All steps we performed before this section were based on the data in FISSSDL. The distribution of sources in the FLUX90-BG90 plot is shown in Fig.~\ref{fig:fluxbg}.

\subsection{Image Inspection}
\label{sec:imageinspection}

\begin{figure*}
    \centering
    \includegraphics[width=0.8\textwidth]{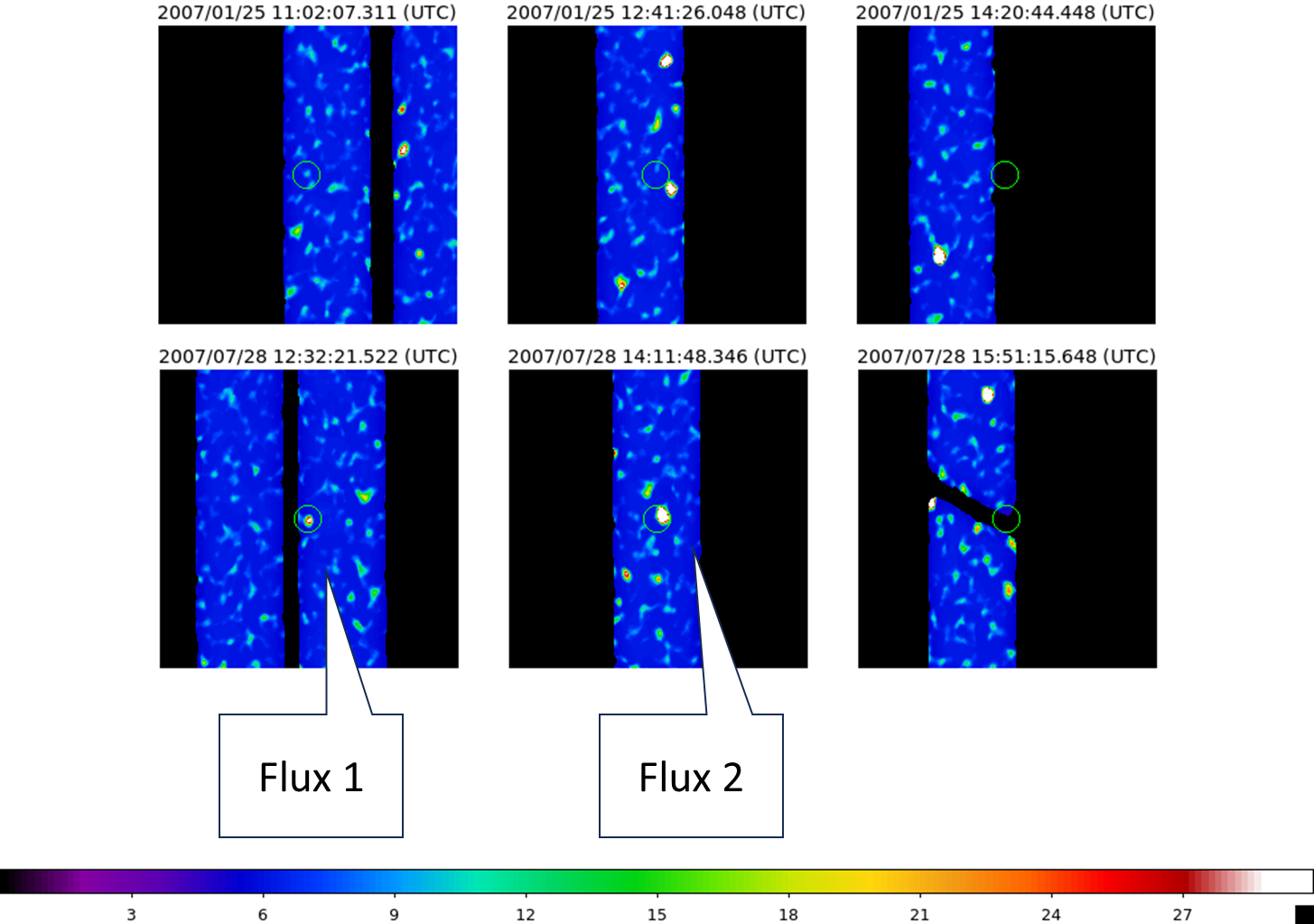}
    \caption{Each scan of FISSSDL J0250422-150114, one of the Planet Nine candidates. The image size is $30'\times30'$, and the green circle is centred at the detection position with an $80''$ radius. The colour represents the likelihood of identifying a point source. The image value is in an arbitrary unit. In the point source extraction, pixels with $\geq$21 are treated as detections at the first step and sent to the confirmation process. FISSSDL J0250422-150114 was detected twice, which is labelled with Flux 1 and 2. The flux values are listed in Table~\ref{tab:candidates_flux}.}
    \label{fig:candi1}
\end{figure*}

We further examine the AKARI ''detection probability maps'' of 393 candidates. The detection probability map is an intermediate data for point source extraction, showing the likelihood of the presence of a point source at the sky position. It is an arbitral unit, but in the FIS, source extraction $>$15 (catalogue unit) is adopted for the WIDE-S threshold for FISBSC. However, during the image inspection, we noticed detections with $<21$ are not reliable. Therefore, we only kept objects with $>$21 (catalogue unit) to guarantee effective detections. 24 out of 393 sources have the second detection with a likelihood between 15 and 21. We removed them since their second detections were not clear, and 369 candidates remained.
Fig.~\ref{fig:candi1} presents the detection probability map for one of the Planet Nine candidates. By analysing the timing of each scan, we can determine whether the candidate is a moving object. Specifically, we identify sources detected at least twice within 24 hours, with no detections at the same location (within AKARI/WIDE-S's beam size of $32''$, see Section~\ref{sec:xmatch}) after six months. Out of 369 candidates, we selected 248 candidates that were clearly detected in all scans within 24 hours.
Out of 248 candidates, 83 uncertain sources without second detection images after a six-month separation period were excluded, as their movement could not be confirmed.  
Ultimately, we selected 165 candidates that were clearly detected in all scans within 24 hours. They are confirmed to have no appearances before six months or disappear after six months.

\begin{figure*}
    \centering
    \includegraphics[width=0.8\textwidth]{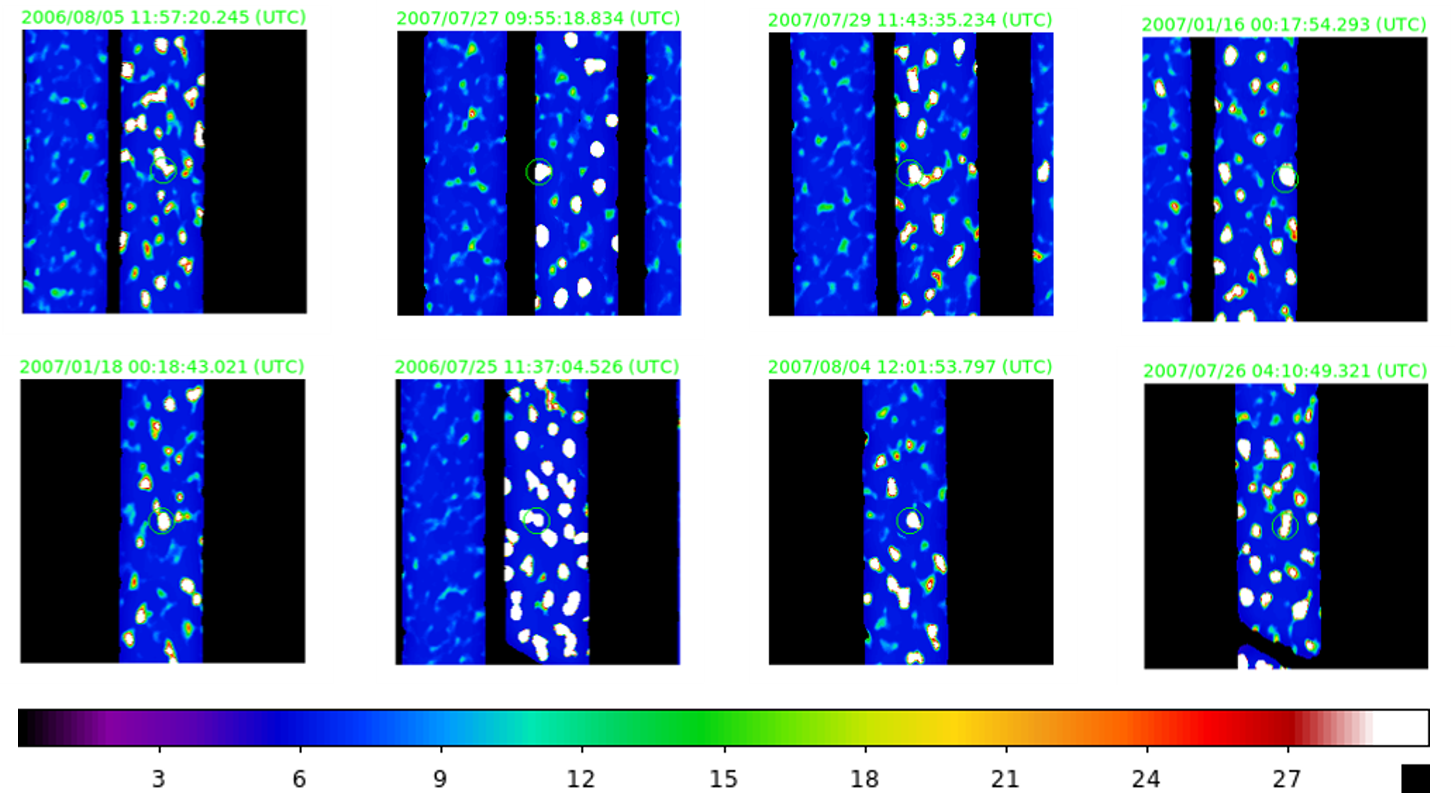}
    \caption{Fake detection caused by CRs when {\it AKARI} pass through SAA. These images are selected from different sources but with similar features. We reject the candidates contaminated by the CRs.}
    \label{fig:saa}
\end{figure*}

During this process, we notice that around 50\% of the "detections" are CRs rather than true objects. Considerable CR hits create a special feature (see Fig.~\ref{fig:saa}) and result in false detection when {\it AKARI} passed the South Atlantic Anomaly (SAA). These sources were rejected in the FISBSC confirmation process. Still, they remained in the FISSSDL because of relaxed conditions (FISBSC requires sources to be detected in at least 3/4 of the total number of scans observed at the position, which is not required for FISSSDL sources). After removing these sources, only 67 candidates are left. 

Out of 67 candidates, 54 candidates' detections are on the edge of the scan data, suggesting they might not be real detections. We removed these 54 candidates, and there are 13 candidates left. It is worth noting that there are 3 possibly fast-moving objects in our 13 Planet Nine candidates. They moved several arcminutes in a few hours, which is too fast to be Planet Nine but could be newly discovered asteroids. After removing these 3 fast-moving objects, we have 10 candidates.

The {\it AKARI} all-sky survey operation lasted about 16 months, so it is possible that {\it AKARI} scanned through the same position with a one-year time separation. In that case, the parallax vanishes, and the proper motion ranges from 1.2 to 6.4 arcminutes. Only 8 out of 10 candidates have detection maps in the subsequent or the preceding year. Four of them have no nearby ($1.2'$-$6.4'$) {\it AKARI} sources, so they are removed from the candidate list. Three of them only matched with monthly confirmed sources, so they were removed. One source has a matched source without monthly confirmation. However, that matched source was detected at the same epoch as our candidate; they are not Planet Nine. Ultimately, we have two Planet Nine candidates that were detected by AKARI in one epoch.
Hereafter, we identify the candidates with their coordinates as FISSSDL Jxxxxxx$\pm$xxxxx and list them in Table~\ref{tab:candidates_flux}.

\begin{figure*}
    \centering
    \includegraphics[width=0.627\textwidth]{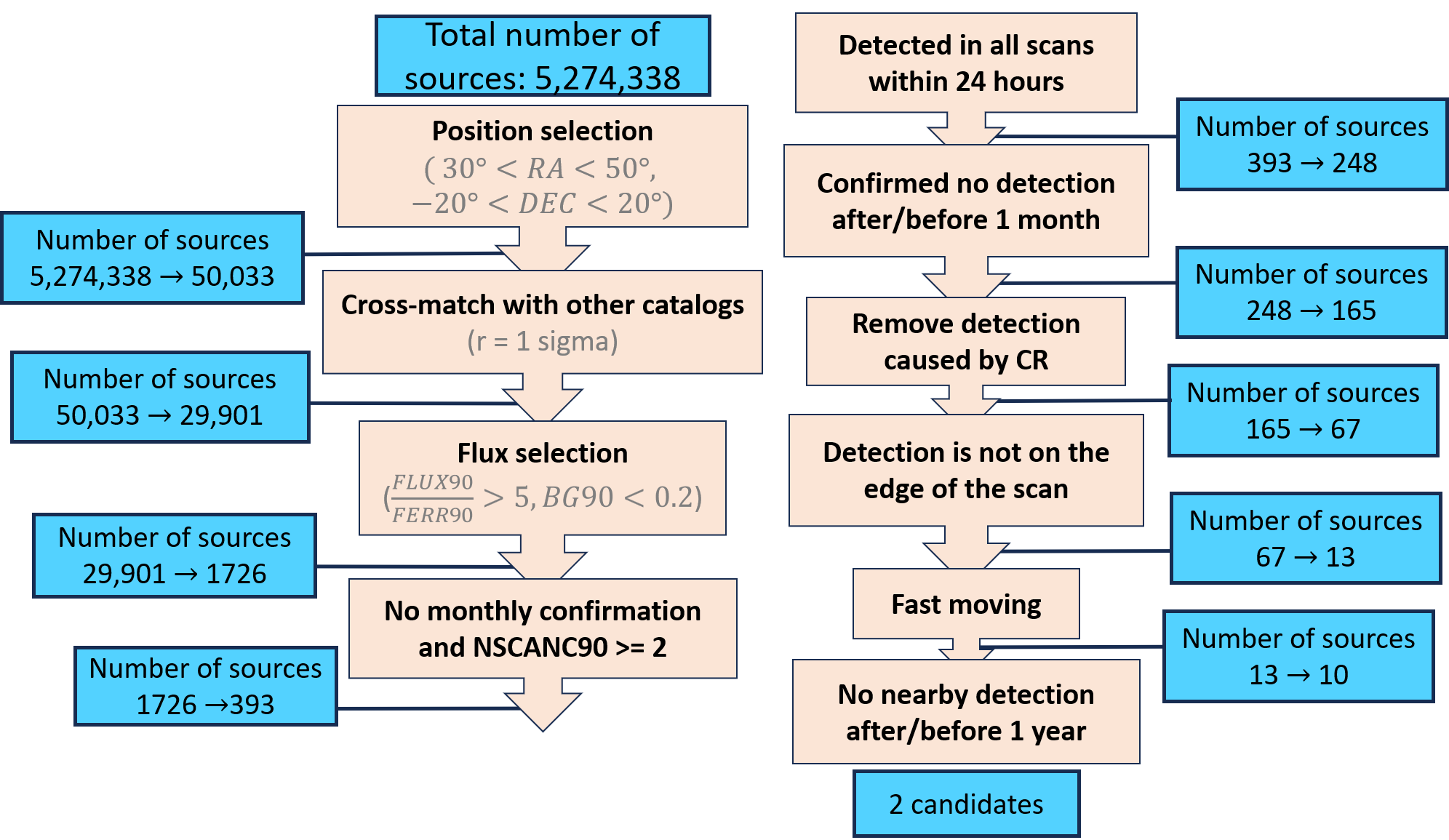}
    \caption{Work flow of this work. Orange blocks are steps we applied to select candidates. Blue blocks show the remaining sources after each step.}
    \label{fig:workflow}
\end{figure*}

A schematic selection process flow chart is shown in Fig.~\ref{fig:workflow}.

\section{Final Planet Nine Candidates}
\label{sec:results}

By identifying moving objects in AKARI FISSSDL and images, we found two Planet Nine candidates.

To analyse the physical properties of Planet Nine candidates, the fluxes of candidates are vital. However, the fluxes recorded in FISSSDL were not always the correct fluxes for these moving objects. In the standard procedure, the flux of a source is measured on the data constructed from all available scans. This process is designed for stable sources but will underestimate the fluxes of moving objects since they are not detected in all scans. We measure each scan's flux of two candidates. The flux values for these candidates vary significantly in each scan, and we have only two detections. Therefore, we present their fluxes for each scan in Table \ref{tab:candidates_flux}. 

\begin{table*}
    \centering
    \begin{tabular}{ccccccccc}
        \hline
        \hline
        Name & \makecell{Flux 1\\(Jy)} & \makecell{Flux 2\\(Jy)} & \makecell{R.A.\\(deg)} & \makecell{Dec.\\(deg)} & \makecell{POSERRMJ\\(arcsec)} & \makecell{POSERRMI\\(arcsec)} & \makecell{POSERRPA\\(deg)} & Detected Epoch\\
        \hline
        FISSSDL J0250422-150114 & 0.61 & 1.62 & 42.676 & -15.021 & 3.5 & 2.3 & 340.3 & 2007/07/28 \\
        \hline
        FISSSDL J0301112-164240 & 1.27 & 0.51 & 45.297 & -16.711 & 3.5 & 2.3 & 340.7 & 2006/07/30 \\
        \hline
        \hline
    \end{tabular}
    \caption{List of two Planet Nine candidates and their 90\,$\mu$m fluxes of each detection. POSERRMJ and POSERRMI are major and minor axes of position error. POSERRPA is the position angle. The epoch of the coordinate system is J2000.}
    \label{tab:candidates_flux}
\end{table*}

During the {\it AKARI}'s all-sky survey, there is a chance that {\it AKARI} detected Planet Nine twice at different positions and times. These two detections might be included in our candidates as two separate candidates. Therefore, we try to find if any pairs of candidates fit the expected Planet Nine parallax among those 83 uncertain sources removed in Section~\ref{sec:detectioin} and the two final candidates. Unfortunately, the separations between each other were all larger than 22.9 arcminutes, which is too large for Planet Nine's predicted parallax (see Fig.~\ref{fig:motion}).

\section{Discussion and Conclusion}
\label{sec:discussion} 

Compared to the combined ZTF \citep{Brown_2022_ztfsearch}, DES \citep{Belyakov_2022_dessearch}, and Pan-STARRS1 \citep{brown2024ps1search} searches, we search for Planet Nine at a larger distance. The combined ZTF, DES, and Pan-STARRS1 survey exclude the situation V magnitude brighter than 21, which roughly corresponds to a distance of 500 au (see Figure 9 of \cite{brown2021orbit}). Our search radius extended to $\sim800$ au if Planet Nine is 53 K.
\cite{C.Sedgwick&S.Serjeant_akari_irassearch} also searched for Planet Nine in the AKARI catalogues. However, they required detections in AKARI FISBSC and IRAS all-sky search and covered the distance range from 700 to 8000 au, which is very different from this work. \cite{Phan2025} conducted a Planet Nine search with a similar scheme to \cite{C.Sedgwick&S.Serjeant_akari_irassearch}, but with more strict criteria on the flux quality and covering a shorter distance range (500 to 700 au). The epochs of IRAS and {\it AKARI} are separated by 23 years, and those two works aim to detect Planet Nine's orbital motion. In contrast, this work is designed to find moving objects that match Planet Nine's half-year parallax.

It is possible that we found a transient event in the AKARI image, rather than a moving object. Here, we estimate the expected events of four types of transient events with a timescale of less than 6 months. To pass our selection criteria and contaminate Planet Nine candidates, the transients need to be brighter than the AKARI/WIDE-S detection limit (0.2 Jy, Figure~\ref{fig:SDLflux}) and be detected twice within one day. The time window for detecting a transient in one year will be $(T-1)/365$, where $T$ is the typical timescale of the transient. We adopt the {\it Planck15} cosmology \citep{Planck2016}, i.e., $\Lambda$ cold dark matter cosmology with ($\Omega_{m}$, $\Omega_{\Lambda}$, $\Omega_{b}$, $h$)=(0.307, 0.693, 0.0486, 0.677) in the volume calculation. The survey volume ($V$) is $\frac{4}{3}\pi \Omega D^3$, where $\Omega$ is the solid angle and $D$ is the maximum luminosity distance to detect a transient.
\begin{itemize}
    \item For core-collapse supernova (CCSN), the event rate is $1.06\times10^{-4} \text{Mpc}^{-3} \text{yr}^{-1}$ \citep{Taylor_2014}. By scaling from the assumed far-infrared flux of around 1 mJy at z=0.083 (377 Mpc) \citep{Perley_2022}, CCSNe need to be closer than $D=27.6$ Mpc to have FLUX90$\geq0.2$ Jy. The volume to the 27.6 Mpc within the survey area is $V=1708\text{ Mpc}^3$. After applying the time window with $T=30$ days \citep{Perley_2022}, we expect only 0.014 CCSNe will be detected within the survey area by the {\it AKARI}. 
    \item For Type Ia supernova (SN Ia), the event rate is $2.43\times10^{-5}\text{Mpc}^{-3}\text{yr}^{-1}$ \citep{Frohmaier2019}. We assume the far-infrared flux is around 10 mJy at 20.23 Mpc \citep{Johansson2013}. The SNe Ia need to be closer than $D=4.5$ Mpc to have FLUX90$\geq0.2$ Jy, where the volume is $V=7.5\text{ Mpc}^3$. After applying the time window with $T=20$ days \citep{Riess_1999}, the expected number of SN Ia detected by {\it AKARI} within the survey area is $1.5\times10^{-5}$.
    \item For Galactic nova, the Galactic nova rate is $\sim30 \text{yr}^{-1}$ \citep{Kawash_2021}. We assume the nova distribution follows the Milky Way stellar mass distribution \citep{Kawash_2021}. The thin disk, thick disk, and stellar halo model are from \cite{Robin2003}. The bulge/bar model is from \cite{Simion2017}. The scale length of the thin and thick disks is 2.5 kpc. $D=15$ kpc is set to be larger than the radius of the Milky Way. The volume to 15 kpc within the survey area is  $V=650\text{ kpc}^3$. Using this model, we scale the nova rate in the Milky Way to $30$ novae per year. After integrating the survey volume, we found that there are only 0.001 novae per year in our survey volume. Since the IR flux of a nova is uncertain, we calculate the expected number of novae in our survey volume. Even assuming {\it AKARI} can detect every nova in the Milky Way, the expected number of detected novae is $7.8\times10^{-5}$ with a time window $T=30$ days \citep{Chomiuk_2021}.
    \item For tidal disruption event (TDE), the event rate is $1.3\times10^{-7} \text{Mpc}^{-3} \text{yr}^{-1}$ \citep{Masterson_2024}, much lower than that of CCSN. Considering TDE is generally fainter than CCSN, TDE events do not contribute to most transient detection.
\end{itemize}
 In conclusion, only $\sim 0.014$ (mostly CCSNe) transient events within one year are expected to be detected by {\it AKARI} in our survey region.

The noise fluctuation could also contribute to fake sources. We assume FISSSDL sources that were detected only once are not real sources. Because Planet Nine selection criteria require at least two hourly detections, we estimate the chance of two bright (FLUX90/FERR90$>5$ and BG90$<0.2$) singly detected sources overlapping. There are 486 singly detected FISSSDL sources with FLUX90/FERR90$>5$ and BG90$<0.2$ in our survey area. The average number of {\it AKARI} scans of these singly detected sources ($N_{scan}$) is 5.3. The mean number of fake sources per scan ($\bar{N}$) is $486/N_{scan}$. 
The density of fake sources ($\rho_f$) is $\bar{N}/A_S$, where $A_S=800 \text{ deg}^2$ is the survey area. The cross-match radius is $32''$ (Section~\ref{sec:xmatch}), so the area of overlapping ($A_O$) is $\pi (32/3600)^2 \text{ deg}^2$. The expected number of overlapping fake sources of two scans will be $\bar{N}\rho_f A_O=0.0025$. With $N_{scan}$ scans, the expected fake is $\bar{N}\rho_fA_O\binom{N_{scan}}{2}=0.029$, where $\binom{N_{scan}}{2} = \frac{N_{scan}!}{N_{scan}!\times(N_{scan}-2)!}$ is the binomial coefficient.

We cross-matched the 3 fast-moving objects (see Section~\ref{sec:imageinspection}) with the Minor Planet Center database\footnote{\url{https://www.minorplanetcenter.net/cgi-bin/checkmp.cgi}} to check if they are known asteroids. All of these 3 fast-moving objects are not known asteroids. However, we found four fast-moving objects in the FISSSDL that are known asteroids. They are Ceres, Patientia, Liguria, and Antiope. They were not on the candidate list since their BG90 are all larger than 0.2.


We select two Planet Nine candidates by identifying moving objects from the AKARI FISSSDL. 
These two objects were detected twice within 24 hours, and not detected after six months.
However, since the fluxes of candidates are degenerated with many parameters, such as distance, mass, temperature, etc., the physical properties of these Planet Nine candidates are hard to constrain. 

To confirm whether any of them is Planet Nine, we need to determine their orbits. However, most of them have only two {\it AKARI} detections, which are insufficient to decide on their orbit. Therefore, follow-up observations are needed. 
For example, future observations can be conducted using the Subaru telescope. Planet Nine's maximum expected angular motion from 2006 to 2024 is $117'$ at 300 au. Many studies have given different predictions by assuming different Planet Nine orbital parameters. \cite{brown2021orbit} estimated Planet Nine has a $16\pm5^{\circ}$ inclination angle while ML17 gave a $30^{\circ}$ inclination angle. To be inclusive, we assume a circular orbit with a minimum radius (300 au) and calculate the maximum displacement.
With an expected R-band magnitude $<26$ \citep{brown2021orbit}, the brightness and motion of Planet Nine are within the capabilities of the Subaru Hyper Suprime-Cam (HSC, \cite{Miyazaki2018}) with a few pointings per target. The HSC on Subaru features a $1.5^{\circ}$ diameter field of view and only needs 503 seconds of exposure time to reach 5-$\sigma$ S/N of a 26 magnitude point source in r2-band.  

The follow-up observation will be essential to verify the Planet Nine hypothesis. The confirmation of Planet Nine and its orbit might be able to explain the orbital clustering of KBOs, which helps us to have a deeper understanding of the solar system's history.

\begin{acknowledgement}
We would like to express our deepest appreciation to the anonymous referee for the comprehensive and thoughtful review of our manuscript. Their detailed examination and insightful suggestions have played a crucial role in refining our work, and the constructive feedback has greatly enhanced the overall quality and clarity of the paper.
TG acknowledges the support of the National Science and Technology Council of Taiwan through grants 108-2628-M-007-004-MY3, 110-2112-M-005-013-MY3, 111-2112-M-007-021, 111-2123-M-001-008-, 112-2112-M-007-013, 112-2123-M-001-004-, 113-2112-M-007-006-, 113-2927-I-007-501-, and 113-2123-M-001-008-. 
TN acknowledges the support by JSPS KAKENHI Grant
Numbers 23H05441 and 23K17695.
TH acknowledges the support of the National Science and Technology Council of Taiwan through grants 110-2112-M-005 -013 -MY3, 113-2112-M-005-009-MY3, 110-2112-M-007-034-, and 113-2123-M-001-008-. 
SH acknowledges the support of the Australian Research Council (ARC) Centre of Excellence (CoE) for Gravitational Wave Discovery (OzGrav) project numbers CE170100004 and CE230100016, and the ARC CoE for All Sky Astrophysics in 3 Dimensions (ASTRO 3D) project number CE170100013.
This research is based on observations with AKARI, a JAXA project with the participation of ESA.
This research has made use of the CDS cross-match service, Strasbourg Astronomical Observatory, France.
This research has made use of the SIMBAD database, operated at CDS, Strasbourg, France.
This publication makes use of data products from the Wide-field Infrared Survey Explorer, which is a joint project of the University of California, Los Angeles, and the Jet Propulsion Laboratory/California Institute of Technology, funded by the National Aeronautics and Space Administration.
This publication makes use of data products from the Two Micron All Sky Survey, which is a joint project of the University of Massachusetts and the Infrared Processing and Analysis Center/California Institute of Technology, funded by the National Aeronautics and Space Administration and the National Science Foundation. This research has made use of services provided by the International Astronomical Union's Minor Planet Center.
\end{acknowledgement}



\printendnotes

\printbibliography

@article{pbh,
  title = {What If Planet 9 Is a Primordial Black Hole?},
  author = {Scholtz, Jakub and Unwin, James},
  journal = {Phys. Rev. Lett.},
  volume = {125},
  issue = {5},
  pages = {051103},
  numpages = {7},
  year = {2020},
  month = {7},
  publisher = {American Physical Society},
  doi = {10.1103/PhysRevLett.125.051103},
  url = {https://link.aps.org/doi/10.1103/PhysRevLett.125.051103}
}

@article{Cowan_2016_cmb,
    doi = {10.3847/2041-8205/822/1/L2},
    url = {https://dx.doi.org/10.3847/2041-8205/822/1/L2},
    year = {2016},
    month = {4},
    publisher = {The American Astronomical Society},
    volume = {822},
    number = {1},
    pages = {L2},
    author = {Nicolas B. Cowan and Gil Holder and Nathan A. Kaib},
    title = {COSMOLOGISTS IN SEARCH OF PLANET NINE: THE CASE FOR CMB EXPERIMENTS},
    journal = {The Astrophysical Journal Letters},
}

@article{Batygin_2016,
    doi = {10.3847/0004-6256/151/2/22},
    url = {https://dx.doi.org/10.3847/0004-6256/151/2/22},
    year = {2016},
    month = {1},
    publisher = {The American Astronomical Society},
    volume = {151},
    number = {2},
    pages = {22},
    author = {Konstantin Batygin and Michael E. Brown},
    title = {EVIDENCE FOR A DISTANT GIANT PLANET IN THE SOLAR SYSTEM},
    journal = {The Astronomical Journal},
}

@article{Brown_2022_ztfsearch,
    doi = {10.3847/1538-3881/ac32dd},
    url = {https://dx.doi.org/10.3847/1538-3881/ac32dd},
    year = {2022},
    month = {1},
    publisher = {The American Astronomical Society},
    volume = {163},
    number = {2},
    pages = {102},
    author = {Michael E. Brown and Konstantin Batygin},
    title = {A Search for Planet Nine using the Zwicky Transient Facility Public Archive},
    journal = {The Astronomical Journal},
}

@article{Batygin_2024,
    doi = {10.3847/2041-8213/ad3cd2},
    url = {https://dx.doi.org/10.3847/2041-8213/ad3cd2},
    year = {2024},
    month = {4},
    publisher = {The American Astronomical Society},
    volume = {966},
    number = {1},
    pages = {L8},
    author = {Konstantin Batygin and Alessandro Morbidelli and Michael E. Brown and David Nesvorný},
    title = {Generation of Low-inclination, Neptune-crossing Trans-Neptunian Objects by Planet Nine},
    journal = {The Astrophysical Journal Letters}
}

@article{Belyakov_2022_dessearch,
    doi = {10.3847/1538-3881/ac5c56},
    url = {https://dx.doi.org/10.3847/1538-3881/ac5c56},
    year = {2022},
    month = {4},
    publisher = {The American Astronomical Society},
    volume = {163},
    number = {5},
    pages = {216},
    author = {Matthew Belyakov and Pedro H. Bernardinelli and Michael E. Brown},
    title = {Limits on the Detection of Planet Nine in the Dark Energy Survey},
    journal = {The Astronomical Journal}
}

@article{IRASsearch,
    author = {Rowan-Robinson, Michael},
    title = "{A search for Planet 9 in the IRAS data}",
    journal = {Monthly Notices of the Royal Astronomical Society},
    volume = {510},
    number = {3},
    pages = {3716-3726},
    year = {2021},
    month = {11},
    issn = {0035-8711},
    doi = {10.1093/mnras/stab3212},
    url = {https://doi.org/10.1093/mnras/stab3212},
    eprint = {https://academic.oup.com/mnras/article-pdf/510/3/3716/45185955/stab3212.pdf},
}

@article{C.Sedgwick&S.Serjeant_akari_irassearch,
    author = {Sedgwick, Chris and Serjeant, Stephen},
    title = "{Searching for giant planets in the outer Solar system with far-infrared all-sky surveys}",
    journal = {Monthly Notices of the Royal Astronomical Society},
    volume = {515},
    number = {4},
    pages = {4828-4837},
    year = {2022},
    month = {07},
    issn = {0035-8711},
    doi = {10.1093/mnras/stac2044},
    url = {https://doi.org/10.1093/mnras/stac2044},
    eprint = {https://academic.oup.com/mnras/article-pdf/515/4/4828/45475230/stac2044.pdf},
}

@ARTICLE{neowisesearch,
       author = {{Meisner}, Aaron M. and {Bromley}, Benjamin C. and {Nugent}, Peter E. and {Schlegel}, David J. and {Kenyon}, Scott J. and {Schlafly}, Edward F. and {Dawson}, Kyle S.},
        title = "{Searching for Planet Nine with Coadded WISE and NEOWISE-Reactivation Images}",
      journal = {\aj},
         year = 2017,
        month = {2},
       volume = {153},
       number = {2},
          eid = {65},
        pages = {65},
          doi = {10.3847/1538-3881/153/2/65},
archivePrefix = {arXiv},
       eprint = {1611.00015},
 primaryClass = {astro-ph.EP},
       adsurl = {https://ui.adsabs.harvard.edu/abs/2017AJ....153...65M}
}

@ARTICLE{wisesearch,
       author = {{Fortney}, Jonathan J. and {Marley}, Mark S. and {Laughlin}, Gregory and {Nettelmann}, Nadine and {Morley}, Caroline V. and {Lupu}, Roxana E. and {Visscher}, Channon and {Jeremic}, Pavle and {Khadder}, Wade G. and {Hargrave}, Mason},
        title = "{The Hunt for Planet Nine: Atmosphere, Spectra, Evolution, and Detectability}",
      journal = {\apjl},
         year = {2016},
        month = {6},
       volume = {824},
       number = {2},
          eid = {L25},
        pages = {L25},
          doi = {10.3847/2041-8205/824/2/L25},
archivePrefix = {arXiv},
       eprint = {1604.07424},
 primaryClass = {astro-ph.EP},
       adsurl = {https://ui.adsabs.harvard.edu/abs/2016ApJ...824L..25F}
}

@article{brown2024ps1search,
      title={A Pan-STARRS1 Search for Planet Nine}, 
      author={Michael E. Brown and Matthew J. Holman and Konstantin Batygin},
      year={2024},
      eprint={2401.17977},
      archivePrefix={arXiv},
      primaryClass={astro-ph.EP},
      url={https://arxiv.org/abs/2401.17977}, 
}

@article{Murakami2007,
author = {Murakami, Hiroshi and Baba, Hajime and Barthel, Peter and Clements, David L. and Cohen, Martin and Doi, Yasuo and Enya, Keigo and Figueredo, Elysandra and Fujishiro, Naofumi and Fujiwara, Hideaki and Fujiwara, Mikio and Garcia-Lario, Pedro and Goto, Tomotsugu and Hasegawa, Sunao and Hibi, Yasunori and Hirao, Takanori and Hiromoto, Norihisa and Hong, Seung Soo and Imai, Koji and Ishigaki, Miho and Ishiguro, Masateru and Ishihara, Daisuke and Ita, Yoshifusa and Jeong, Woong-Seob and Jeong, Kyung Sook and Kaneda, Hidehiro and Kataza, Hirokazu and Kawada, Mitsunobu and Kawai, Toshihide and Kawamura, Akiko and Kessler, Martin F. and Kester, Do and Kii, Tsuneo and Kim, Dong Chan and Kim, Woojung and Kobayashi, Hisato and Koo, Bon Chul and Kwon, Suk Minn and Lee, Hyung Mok and Lorente, Rosario and Makiuti, Sin’itirou and Matsuhara, Hideo and Matsumoto, Toshio and Matsuo, Hiroshi and Matsuura, Shuji and MÜller, Thomas G. and Murakami, Noriko and Nagata, Hirohisa and Nakagawa, Takao and Naoi, Takahiro and Narita, Masanao and Noda, Manabu and Oh, Sang Hoon and Ohnishi, Akira and Ohyama, Youichi and Okada, Yoko and Okuda, Haruyuki and Oliver, Sebastian and Onaka, Takashi and Ootsubo, Takafumi and Oyabu, Shinki and Pak, Soojong and Park, Yong-Sun and Pearson, Chris P. and Rowan-Robinson, Michael and Saito, Toshinobu and Sakon, Itsuki and Salama, Alberto and Sato, Shinji and Savage, Richard S. and Serjeant, Stephen and Shibai, Hiroshi and Shirahata, Mai and Sohn, Jungjoo and Suzuki, Toyoaki and Takagi, Toshinobu and Takahashi, Hidenori and TanabÉ, Toshihiko and Takeuchi, Tsutomu T. and Takita, Satoshi and Thomson, Matthew and Uemizu, Kazunori and Ueno, Munetaka and Usui, Fumihiko and Verdugo, Eva and Wada, Takehiko and Wang, Lingyu and Watabe, Toyoki and Watarai, Hidenori and White, Glenn J. and Yamamura, Issei and Yamauchi, Chisato and Yasuda, Akiko},
title = "{The Infrared Astronomical Mission AKARI*}",
journal = {Publications of the Astronomical Society of Japan},
volume = {59},
number = {sp2},
pages = {S369-S376},
year = {2007},
month = {10},
issn = {0004-6264},
doi = {10.1093/pasj/59.sp2.S369},
url = {https://doi.org/10.1093/pasj/59.sp2.S369},
eprint = {https://academic.oup.com/pasj/article-pdf/59/sp2/S369/54713437/pasj\_59\_sp2\_s369.pdf},
}

@article{Millholland_2017,
doi = {10.3847/1538-3881/153/3/91},
url = {https://dx.doi.org/10.3847/1538-3881/153/3/91},
year = {2017},
month = {2},
publisher = {The American Astronomical Society},
volume = {153},
number = {3},
pages = {91},
author = {Sarah Millholland and Gregory Laughlin},
title = {Constraints on Planet Nine’s Orbit and Sky Position within a Framework of Mean-motion Resonances},
journal = {The Astronomical Journal},
}

@article{brown2021orbit,
    doi = {10.3847/1538-3881/ac2056},
    url = {https://dx.doi.org/10.3847/1538-3881/ac2056},
    year = {2021},
    month = {10},
    publisher = {The American Astronomical Society},
    volume = {162},
    number = {5},
    pages = {219},
    author = {Michael E. Brown and Konstantin Batygin},
    title = {The Orbit of Planet Nine},
    journal = {The Astronomical Journal}
}

@article{Brown_2019_bias,
    doi = {10.3847/1538-3881/aaf051},
    url = {https://dx.doi.org/10.3847/1538-3881/aaf051},
    year = {2019},
    month = {1},
    publisher = {The American Astronomical Society},
    volume = {157},
    number = {2},
    pages = {62},
    author = {Michael E. Brown and Konstantin Batygin},
    title = {Orbital Clustering in the Distant Solar System},
    journal = {The Astronomical Journal}
}

@ARTICLE{2mass,
       author = {{Skrutskie}, M.~F. and {Cutri}, R.~M. and {Stiening}, R. and {Weinberg}, M.~D. and {Schneider}, S. and {Carpenter}, J.~M. and {Beichman}, C. and {Capps}, R. and {Chester}, T. and {Elias}, J. and {Huchra}, J. and {Liebert}, J. and {Lonsdale}, C. and {Monet}, D.~G. and {Price}, S. and {Seitzer}, P. and {Jarrett}, T. and {Kirkpatrick}, J.~D. and {Gizis}, J.~E. and {Howard}, E. and {Evans}, T. and {Fowler}, J. and {Fullmer}, L. and {Hurt}, R. and {Light}, R. and {Kopan}, E.~L. and {Marsh}, K.~A. and {McCallon}, H.~L. and {Tam}, R. and {Van Dyk}, S. and {Wheelock}, S.},
        title = "{The Two Micron All Sky Survey (2MASS)}",
      journal = {\aj},
         year = {2006},
        month = {2},
       volume = {131},
       number = {2},
        pages = {1163-1183},
          doi = {10.1086/498708},
       adsurl = {https://ui.adsabs.harvard.edu/abs/2006AJ....131.1163S}
}

@ARTICLE{ps1,
       author = {{Chambers}, K.~C. and {Magnier}, E.~A. and {Metcalfe}, N. and {Flewelling}, H.~A. and {Huber}, M.~E. and {Waters}, C.~Z. and {Denneau}, L. and {Draper}, P.~W. and {Farrow}, D. and {Finkbeiner}, D.~P. and {Holmberg}, C. and {Koppenhoefer}, J. and {Price}, P.~A. and {Rest}, A. and {Saglia}, R.~P. and {Schlafly}, E.~F. and {Smartt}, S.~J. and {Sweeney}, W. and {Wainscoat}, R.~J. and {Burgett}, W.~S. and {Chastel}, S. and {Grav}, T. and {Heasley}, J.~N. and {Hodapp}, K.~W. and {Jedicke}, R. and {Kaiser}, N. and {Kudritzki}, R. -P. and {Luppino}, G.~A. and {Lupton}, R.~H. and {Monet}, D.~G. and {Morgan}, J.~S. and {Onaka}, P.~M. and {Shiao}, B. and {Stubbs}, C.~W. and {Tonry}, J.~L. and {White}, R. and {Ba{\~n}ados}, E. and {Bell}, E.~F. and {Bender}, R. and {Bernard}, E.~J. and {Boegner}, M. and {Boffi}, F. and {Botticella}, M.~T. and {Calamida}, A. and {Casertano}, S. and {Chen}, W. -P. and {Chen}, X. and {Cole}, S. and {Deacon}, N. and {Frenk}, C. and {Fitzsimmons}, A. and {Gezari}, S. and {Gibbs}, V. and {Goessl}, C. and {Goggia}, T. and {Gourgue}, R. and {Goldman}, B. and {Grant}, P. and {Grebel}, E.~K. and {Hambly}, N.~C. and {Hasinger}, G. and {Heavens}, A.~F. and {Heckman}, T.~M. and {Henderson}, R. and {Henning}, T. and {Holman}, M. and {Hopp}, U. and {Ip}, W. -H. and {Isani}, S. and {Jackson}, M. and {Keyes}, C.~D. and {Koekemoer}, A.~M. and {Kotak}, R. and {Le}, D. and {Liska}, D. and {Long}, K.~S. and {Lucey}, J.~R. and {Liu}, M. and {Martin}, N.~F. and {Masci}, G. and {McLean}, B. and {Mindel}, E. and {Misra}, P. and {Morganson}, E. and {Murphy}, D.~N.~A. and {Obaika}, A. and {Narayan}, G. and {Nieto-Santisteban}, M.~A. and {Norberg}, P. and {Peacock}, J.~A. and {Pier}, E.~A. and {Postman}, M. and {Primak}, N. and {Rae}, C. and {Rai}, A. and {Riess}, A. and {Riffeser}, A. and {Rix}, H.~W. and {R{\"o}ser}, S. and {Russel}, R. and {Rutz}, L. and {Schilbach}, E. and {Schultz}, A.~S.~B. and {Scolnic}, D. and {Strolger}, L. and {Szalay}, A. and {Seitz}, S. and {Small}, E. and {Smith}, K.~W. and {Soderblom}, D.~R. and {Taylor}, P. and {Thomson}, R. and {Taylor}, A.~N. and {Thakar}, A.~R. and {Thiel}, J. and {Thilker}, D. and {Unger}, D. and {Urata}, Y. and {Valenti}, J. and {Wagner}, J. and {Walder}, T. and {Walter}, F. and {Watters}, S.~P. and {Werner}, S. and {Wood-Vasey}, W.~M. and {Wyse}, R.},
        title = "{The Pan-STARRS1 Surveys}",
      journal = {arXiv e-prints},
         year = {2016},
        month = {12},
          eid = {arXiv:1612.05560},
        pages = {arXiv:1612.05560},
          doi = {10.48550/arXiv.1612.05560},
archivePrefix = {arXiv},
       eprint = {1612.05560},
 primaryClass = {astro-ph.IM},
       adsurl = {https://ui.adsabs.harvard.edu/abs/2016arXiv161205560C}
}

@ARTICLE{wise,
       author = {{Wright}, Edward L. and {Eisenhardt}, Peter R.~M. and {Mainzer}, Amy K. and {Ressler}, Michael E. and {Cutri}, Roc M. and {Jarrett}, Thomas and {Kirkpatrick}, J. Davy and {Padgett}, Deborah and {McMillan}, Robert S. and {Skrutskie}, Michael and {Stanford}, S.~A. and {Cohen}, Martin and {Walker}, Russell G. and {Mather}, John C. and {Leisawitz}, David and {Gautier}, Thomas N., III and {McLean}, Ian and {Benford}, Dominic and {Lonsdale}, Carol J. and {Blain}, Andrew and {Mendez}, Bryan and {Irace}, William R. and {Duval}, Valerie and {Liu}, Fengchuan and {Royer}, Don and {Heinrichsen}, Ingolf and {Howard}, Joan and {Shannon}, Mark and {Kendall}, Martha and {Walsh}, Amy L. and {Larsen}, Mark and {Cardon}, Joel G. and {Schick}, Scott and {Schwalm}, Mark and {Abid}, Mohamed and {Fabinsky}, Beth and {Naes}, Larry and {Tsai}, Chao-Wei},
        title = "{The Wide-field Infrared Survey Explorer (WISE): Mission Description and Initial On-orbit Performance}",
      journal = {\aj},
         year = {2010},
        month = {12},
       volume = {140},
       number = {6},
        pages = {1868-1881},
          doi = {10.1088/0004-6256/140/6/1868},
archivePrefix = {arXiv},
       eprint = {1008.0031},
 primaryClass = {astro-ph.IM},
       adsurl = {https://ui.adsabs.harvard.edu/abs/2010AJ....140.1868W}
}

@ARTICLE{neowise,
       author = {{Mainzer}, A. and {Bauer}, J. and {Grav}, T. and {Masiero}, J. and {Cutri}, R.~M. and {Dailey}, J. and {Eisenhardt}, P. and {McMillan}, R.~S. and {Wright}, E. and {Walker}, R. and {Jedicke}, R. and {Spahr}, T. and {Tholen}, D. and {Alles}, R. and {Beck}, R. and {Brandenburg}, H. and {Conrow}, T. and {Evans}, T. and {Fowler}, J. and {Jarrett}, T. and {Marsh}, K. and {Masci}, F. and {McCallon}, H. and {Wheelock}, S. and {Wittman}, M. and {Wyatt}, P. and {DeBaun}, E. and {Elliott}, G. and {Elsbury}, D. and {Gautier}, T., IV and {Gomillion}, S. and {Leisawitz}, D. and {Maleszewski}, C. and {Micheli}, M. and {Wilkins}, A.},
        title = "{Preliminary Results from NEOWISE: An Enhancement to the Wide-field Infrared Survey Explorer for Solar System Science}",
      journal = {\apj},
         year = {2011},
        month = {4},
       volume = {731},
       number = {1},
          eid = {53},
        pages = {53},
          doi = {10.1088/0004-637X/731/1/53},
archivePrefix = {arXiv},
       eprint = {1102.1996},
 primaryClass = {astro-ph.EP},
       adsurl = {https://ui.adsabs.harvard.edu/abs/2011ApJ...731...53M}
}

@article{catwise,
doi = {10.3847/1538-4365/abd805},
url = {https://dx.doi.org/10.3847/1538-4365/abd805},
year = {2021},
month = {2},
publisher = {The American Astronomical Society},
volume = {253},
number = {1},
pages = {8},
author = {Federico Marocco and Peter R. M. Eisenhardt and John W. Fowler and J. Davy Kirkpatrick and Aaron M. Meisner and Edward F. Schlafly and S. A. Stanford and Nelson Garcia and Dan Caselden and Michael C. Cushing and Roc M. Cutri and Jacqueline K. Faherty and Christopher R. Gelino and Anthony H. Gonzalez and Thomas H. Jarrett and Renata Koontz and Amanda Mainzer and Elijah J. Marchese and Bahram Mobasher and David J. Schlegel and Daniel Stern and Harry I. Teplitz and Edward L. Wright},
title = {The CatWISE2020 Catalog},
journal = {The Astrophysical Journal Supplement Series},}

@article{unwise,
doi = {10.3847/1538-4365/aafbea},
url = {https://dx.doi.org/10.3847/1538-4365/aafbea},
year = {2019},
month = {2},
publisher = {The American Astronomical Society},
volume = {240},
number = {2},
pages = {30},
author = {Edward F. Schlafly and Aaron M. Meisner and Gregory M. Green},
title = {The unWISE Catalog: Two Billion Infrared Sources from Five Years of WISE Imaging},
journal = {The Astrophysical Journal Supplement Series},
}

@ARTICLE{simbad,
       author = {{Wenger}, M. and {Ochsenbein}, F. and {Egret}, D. and {Dubois}, P. and {Bonnarel}, F. and {Borde}, S. and {Genova}, F. and {Jasniewicz}, G. and {Lalo{\"e}}, S. and {Lesteven}, S. and {Monier}, R.},
        title = "{The SIMBAD astronomical database. The CDS reference database for astronomical objects}",
      journal = {\aaps},
         year = {2000},
        month = {4},
       volume = {143},
        pages = {9-22},
          doi = {10.1051/aas:2000332},
archivePrefix = {arXiv},
       eprint = {astro-ph/0002110},
 primaryClass = {astro-ph},
       adsurl = {https://ui.adsabs.harvard.edu/abs/2000A&AS..143....9W}
}

@ARTICLE{sdss,
       author = {{Ahumada}, Romina and {Allende Prieto}, Carlos and {Almeida}, Andr{\'e}s and {Anders}, Friedrich and {Anderson}, Scott F. and {Andrews}, Brett H. and {Anguiano}, Borja and {Arcodia}, Riccardo and {Armengaud}, Eric and {Aubert}, Marie and {Avila}, Santiago and {Avila-Reese}, Vladimir and {Badenes}, Carles and {Balland}, Christophe and {Barger}, Kat and {Barrera-Ballesteros}, Jorge K. and {Basu}, Sarbani and {Bautista}, Julian and {Beaton}, Rachael L. and {Beers}, Timothy C. and {Benavides}, B. Izamar T. and {Bender}, Chad F. and {Bernardi}, Mariangela and {Bershady}, Matthew and {Beutler}, Florian and {Bidin}, Christian Moni and {Bird}, Jonathan and {Bizyaev}, Dmitry and {Blanc}, Guillermo A. and {Blanton}, Michael R. and {Boquien}, M{\'e}d{\'e}ric and {Borissova}, Jura and {Bovy}, Jo and {Brandt}, W.~N. and {Brinkmann}, Jonathan and {Brownstein}, Joel R. and {Bundy}, Kevin and {Bureau}, Martin and {Burgasser}, Adam and {Burtin}, Etienne and {Cano-D{\'\i}az}, Mariana and {Capasso}, Raffaella and {Cappellari}, Michele and {Carrera}, Ricardo and {Chabanier}, Sol{\`e}ne and {Chaplin}, William and {Chapman}, Michael and {Cherinka}, Brian and {Chiappini}, Cristina and {Doohyun Choi}, Peter and {Chojnowski}, S. Drew and {Chung}, Haeun and {Clerc}, Nicolas and {Coffey}, Damien and {Comerford}, Julia M. and {Comparat}, Johan and {da Costa}, Luiz and {Cousinou}, Marie-Claude and {Covey}, Kevin and {Crane}, Jeffrey D. and {Cunha}, Katia and {Ilha}, Gabriele da Silva and {Dai}, Yu Sophia and {Damsted}, Sanna B. and {Darling}, Jeremy and {Davidson}, James W., Jr. and {Davies}, Roger and {Dawson}, Kyle and {De}, Nikhil and {de la Macorra}, Axel and {De Lee}, Nathan and {Queiroz}, Anna B{\'a}rbara de Andrade and {Deconto Machado}, Alice and {de la Torre}, Sylvain and {Dell'Agli}, Flavia and {du Mas des Bourboux}, H{\'e}lion and {Diamond-Stanic}, Aleksandar M. and {Dillon}, Sean and {Donor}, John and {Drory}, Niv and {Duckworth}, Chris and {Dwelly}, Tom and {Ebelke}, Garrett and {Eftekharzadeh}, Sarah and {Davis Eigenbrot}, Arthur and {Elsworth}, Yvonne P. and {Eracleous}, Mike and {Erfanianfar}, Ghazaleh and {Escoffier}, Stephanie and {Fan}, Xiaohui and {Farr}, Emily and {Fern{\'a}ndez-Trincado}, Jos{\'e} G. and {Feuillet}, Diane and {Finoguenov}, Alexis and {Fofie}, Patricia and {Fraser-McKelvie}, Amelia and {Frinchaboy}, Peter M. and {Fromenteau}, Sebastien and {Fu}, Hai and {Galbany}, Llu{\'\i}s and {Garcia}, Rafael A. and {Garc{\'\i}a-Hern{\'a}ndez}, D.~A. and {Garma Oehmichen}, Luis Alberto and {Ge}, Junqiang and {Geimba Maia}, Marcio Antonio and {Geisler}, Doug and {Gelfand}, Joseph and {Goddy}, Julian and {Gonzalez-Perez}, Violeta and {Grabowski}, Kathleen and {Green}, Paul and {Grier}, Catherine J. and {Guo}, Hong and {Guy}, Julien and {Harding}, Paul and {Hasselquist}, Sten and {Hawken}, Adam James and {Hayes}, Christian R. and {Hearty}, Fred and {Hekker}, S. and {Hogg}, David W. and {Holtzman}, Jon A. and {Horta}, Danny and {Hou}, Jiamin and {Hsieh}, Bau-Ching and {Huber}, Daniel and {Hunt}, Jason A.~S. and {Ider Chitham}, J. and {Imig}, Julie and {Jaber}, Mariana and {Jimenez Angel}, Camilo Eduardo and {Johnson}, Jennifer A. and {Jones}, Amy M. and {J{\"o}nsson}, Henrik and {Jullo}, Eric and {Kim}, Yerim and {Kinemuchi}, Karen and {Kirkpatrick}, Charles C., IV and {Kite}, George W. and {Klaene}, Mark and {Kneib}, Jean-Paul and {Kollmeier}, Juna A. and {Kong}, Hui and {Kounkel}, Marina and {Krishnarao}, Dhanesh and {Lacerna}, Ivan and {Lan}, Ting-Wen and {Lane}, Richard R. and {Law}, David R. and {Le Goff}, Jean-Marc and {Leung}, Henry W. and {Lewis}, Hannah and {Li}, Cheng and {Lian}, Jianhui and {Lin}, Lihwai and {Long}, Dan and {Longa-Pe{\~n}a}, Pen{\'e}lope and {Lundgren}, Britt and {Lyke}, Brad W. and {Mackereth}, J. Ted and {MacLeod}, Chelsea L. and {Majewski}, Steven R. and {Manchado}, Arturo and {Maraston}, Claudia and {Martini}, Paul and {Masseron}, Thomas and {Masters}, Karen L. and {Mathur}, Savita and {McDermid}, Richard M. and {Merloni}, Andrea and {Merrifield}, Michael and {M{\'e}sz{\'a}ros}, Szabolcs and {Miglio}, Andrea and {Minniti}, Dante and {Minsley}, Rebecca and {Miyaji}, Takamitsu and {Mohammad}, Faizan Gohar and {Mosser}, Benoit and {Mueller}, Eva-Maria and {Muna}, Demitri and {Mu{\~n}oz-Guti{\'e}rrez}, Andrea and {Myers}, Adam D. and {Nadathur}, Seshadri and {Nair}, Preethi and {Nandra}, Kirpal and {Correa do Nascimento}, Janaina and {Nevin}, Rebecca Jean and {Newman}, Jeffrey A. and {Nidever}, David L. and {Nitschelm}, Christian and {Noterdaeme}, Pasquier and {O'Connell}, Julia E. and {Olmstead}, Matthew D. and {Oravetz}, Daniel and {Oravetz}, Audrey and {Osorio}, Yeisson and {Pace}, Zachary J. and {Padilla}, Nelson and {Palanque-Delabrouille}, Nathalie and {Palicio}, Pedro A. and {Pan}, Hsi-An and {Pan}, Kaike and {Parker}, James and {Paviot}, Romain and {Peirani}, Sebastien and {Ram{\'r}ez}, Karla Pe{\~n}a and {Penny}, Samantha and {Percival}, Will J. and {Perez-Fournon}, Ismael and {P{\'e}rez-R{\`a}fols}, Ignasi and {Petitjean}, Patrick and {Pieri}, Matthew M. and {Pinsonneault}, Marc and {Poovelil}, Vijith Jacob and {Povick}, Joshua Tyler and {Prakash}, Abhishek and {Price-Whelan}, Adrian M. and {Raddick}, M. Jordan and {Raichoor}, Anand and {Ray}, Amy and {Rembold}, Sandro Barboza and {Rezaie}, Mehdi and {Riffel}, Rogemar A. and {Riffel}, Rog{\'e}rio and {Rix}, Hans-Walter and {Robin}, Annie C. and {Roman-Lopes}, A. and {Rom{\'a}n-Z{\'u}{\~n}iga}, Carlos and {Rose}, Benjamin and {Ross}, Ashley J. and {Rossi}, Graziano and {Rowlands}, Kate and {Rubin}, Kate H.~R. and {Salvato}, Mara and {S{\'a}nchez}, Ariel G. and {S{\'a}nchez-Menguiano}, Laura and {S{\'a}nchez-Gallego}, Jos{\'e} R. and {Sayres}, Conor and {Schaefer}, Adam and {Schiavon}, Ricardo P. and {Schimoia}, Jaderson S. and {Schlafly}, Edward and {Schlegel}, David and {Schneider}, Donald P. and {Schultheis}, Mathias and {Schwope}, Axel and {Seo}, Hee-Jong and {Serenelli}, Aldo and {Shafieloo}, Arman and {Shamsi}, Shoaib Jamal and {Shao}, Zhengyi and {Shen}, Shiyin and {Shetrone}, Matthew and {Shirley}, Raphael and {Silva Aguirre}, V{\'\i}ctor and {Simon}, Joshua D. and {Skrutskie}, M.~F. and {Slosar}, An{\v{z}}e and {Smethurst}, Rebecca and {Sobeck}, Jennifer and {Sodi}, Bernardo Cervantes and {Souto}, Diogo and {Stark}, David V. and {Stassun}, Keivan G. and {Steinmetz}, Matthias and {Stello}, Dennis and {Stermer}, Julianna and {Storchi-Bergmann}, Thaisa and {Streblyanska}, Alina and {Stringfellow}, Guy S. and {Stutz}, Amelia and {Su{\'a}rez}, Genaro and {Sun}, Jing and {Taghizadeh-Popp}, Manuchehr and {Talbot}, Michael S. and {Tayar}, Jamie and {Thakar}, Aniruddha R. and {Theriault}, Riley and {Thomas}, Daniel and {Thomas}, Zak C. and {Tinker}, Jeremy and {Tojeiro}, Rita and {Toledo}, Hector Hernandez and {Tremonti}, Christy A. and {Troup}, Nicholas W. and {Tuttle}, Sarah and {Unda-Sanzana}, Eduardo and {Valentini}, Marica and {Vargas-Gonz{\'a}lez}, Jaime and {Vargas-Maga{\~n}a}, Mariana and {V{\'a}zquez-Mata}, Jose Antonio and {Vivek}, M. and {Wake}, David and {Wang}, Yuting and {Weaver}, Benjamin Alan and {Weijmans}, Anne-Marie and {Wild}, Vivienne and {Wilson}, John C. and {Wilson}, Robert F. and {Wolthuis}, Nathan and {Wood-Vasey}, W.~M. and {Yan}, Renbin and {Yang}, Meng and {Y{\`e}che}, Christophe and {Zamora}, Olga and {Zarrouk}, Pauline and {Zasowski}, Gail and {Zhang}, Kai and {Zhao}, Cheng and {Zhao}, Gongbo and {Zheng}, Zheng and {Zheng}, Zheng and {Zhu}, Guangtun and {Zou}, Hu},
        title = "{The 16th Data Release of the Sloan Digital Sky Surveys: First Release from the APOGEE-2 Southern Survey and Full Release of eBOSS Spectra}",
      journal = {\apjs},
         year = {2020},
        month = {7},
       volume = {249},
       number = {1},
          eid = {3},
        pages = {3},
          doi = {10.3847/1538-4365/ab929e},
archivePrefix = {arXiv},
       eprint = {1912.02905},
 primaryClass = {astro-ph.GA},
       adsurl = {https://ui.adsabs.harvard.edu/abs/2020ApJS..249....3A}
}

@article{nomad,
       author = {{Zacharias}, N. and {Monet}, D.~G. and {Levine}, S.~E. and {Urban}, S.~E. and {Gaume}, R. and {Wycoff}, G.~L.},
        title = "{The Naval Observatory Merged Astrometric Dataset (NOMAD)}",
    booktitle = {American Astronomical Society Meeting Abstracts},
         year = {2004},
       series = {American Astronomical Society Meeting Abstracts},
       volume = {205},
        month = {12},
          eid = {48.15},
        pages = {48.15},
        url = {https://ui.adsabs.harvard.edu/abs/2004AAS...205.4815Z}
}

@article{Kawada2007,
    author = {Kawada, Mitsunobu and Baba, Hajime and Barthel, Peter D. and Clements, David and Cohen, Martin and Doi, Yasuo and Figueredo, Elysandra and Fujiwara, Mikio and Goto, Tomotsugu and Hasegawa, Sunao and Hibi, Yasunori and Hirao, Takanori and Hiromoto, Norihisa and Jeong, Woong-Seob and Kaneda, Hidehiro and Kawai, Toshihide and Kawamura, Akiko and Kester, Do and Kii, Tsuneo and Kobayashi, Hisato and Kwon, Suk Minn and Lee, Hyung Mok and Makiuti, Sin’itirou and Matsuo, Hiroshi and Matsuura, Shuji and MÜller, Thomas G. and Murakami, Noriko and Nagata, Hirohisa and Nakagawa, Takao and Narita, Masanao and Noda, Manabu and Oh, Sang Hoon and Okada, Yoko and Okuda, Haruyuki and Oliver, Sebastian and Ootsubo, Takafumi and Pak, Soojong and Park, Yong-Sun and Pearson, Chris P. and Rowan-Robinson, Michael and Saito, Toshinobu and Salama, Alberto and Sato, Shinji and Savage, Richard S. and Serjeant, Stephen and Shibai, Hiroshi and Shirahata, Mai and Sohn, Jungjoo and Suzuki, Toyoaki and Takagi, Toshinobu and Takahashi, Hidenori and Thomson, Matthew and Usui, Fumihiko and Verdugo, Eva and Watabe, Toyoki and White, Glenn J. and Wang, Lingyu and Yamamura, Issei and Yamauchi, Chisato and Yasuda, Akiko},
    title = "{The Far-Infrared Surveyor (FIS) for AKARI*}",
    journal = {Publications of the Astronomical Society of Japan},
    volume = {59},
    number = {sp2},
    pages = {S389-S400},
    year = {2007},
    month = {10},
    issn = {0004-6264},
    doi = {10.1093/pasj/59.sp2.S389},
    url = {https://doi.org/10.1093/pasj/59.sp2.S389},
    eprint = {https://academic.oup.com/pasj/article-pdf/59/sp2/S389/54713554/pasj\_59\_sp2\_s389.pdf},
}

@article{Onaka2007,
    author = {Onaka, Takashi and Matsuhara, Hideo and Wada, Takehiko and Fujishiro, Naofumi and Fujiwara, Hideaki and Ishigaki, Miho and Ishihara, Daisuke and Ita, Yoshifusa and Kataza, Hirokazu and Kim, Woojung and Matsumoto, Toshio and Murakami, Hiroshi and Ohyama, Youichi and Oyabu, Shinki and Sakon, Itsuki and TanabÉ, Toshihiko and Takagi, Toshinobu and Uemizu, Kazunori and Ueno, Munetaka and Usui, Fumio and Watarai, Hidenori and Cohen, Martin and Enya, Keigo and Ootsubo, Takafumi and Pearson, Chris P. and Takeyama, Norihide and Yamamuro, Tomoyasu and Ikeda, Yuji},
    title = "{The Infrared Camera (IRC) for AKARI–Design and Imaging Performance}",
    journal = {Publications of the Astronomical Society of Japan},
    volume = {59},
    number = {sp2},
    pages = {S401-S410},
    year = {2007},
    month = {10},
    issn = {0004-6264},
    doi = {10.1093/pasj/59.sp2.S401},
    url = {https://doi.org/10.1093/pasj/59.sp2.S401},
    eprint = {https://academic.oup.com/pasj/article-pdf/59/sp2/S401/54713463/pasj\_59\_sp2\_s401.pdf},
}

@ARTICLE{Shankman2017,
       author = {{Shankman}, Cory and {Kavelaars}, J.~J. and {Bannister}, Michele T. and {Gladman}, Brett J. and {Lawler}, Samantha M. and {Chen}, Ying-Tung and {Jakubik}, Marian and {Kaib}, Nathan and {Alexandersen}, Mike and {Gwyn}, Stephen D.~J. and {Petit}, Jean-Marc and {Volk}, Kathryn},
        title = "{OSSOS. VI. Striking Biases in the Detection of Large Semimajor Axis Trans-Neptunian Objects}",
      journal = {\aj},
         year = 2017,
        month = aug,
       volume = {154},
       number = {2},
          eid = {50},
        pages = {50},
          doi = {10.3847/1538-3881/aa7aed},
archivePrefix = {arXiv},
       eprint = {1706.05348},
 primaryClass = {astro-ph.EP},
       adsurl = {https://ui.adsabs.harvard.edu/abs/2017AJ....154...50S}
}

@ARTICLE{Bernardinelli2020,
       author = {{Bernardinelli}, Pedro H. and {Bernstein}, Gary M. and {Sako}, Masao and {Hamilton}, Stephanie and {Gerdes}, David W. and {Adams}, Fred C. and {Saunders}, William R. and {Aguena}, M. and {Allam}, S. and {Avila}, S. and {Brooks}, D. and {Diehl}, H.~T. and {Doel}, P. and {Everett}, S. and {Garc{\'\i}a-Bellido}, J. and {Gaztanaga}, E. and {Gruendl}, R.~A. and {Honscheid}, K. and {Ogando}, R.~L.~C. and {Palmese}, A. and {Tucker}, D.~L. and {Walker}, A.~R. and {Wester}, W. and {(The DES Collaboration)}},
        title = "{Testing the Isotropy of the Dark Energy Survey's Extreme Trans-Neptunian Objects}",
      journal = {\psj},
         year = 2020,
        month = sep,
       volume = {1},
       number = {2},
          eid = {28},
        pages = {28},
          doi = {10.3847/PSJ/ab9d80},
archivePrefix = {arXiv},
       eprint = {2003.08901},
 primaryClass = {astro-ph.EP},
       adsurl = {https://ui.adsabs.harvard.edu/abs/2020PSJ.....1...28B}
}

@ARTICLE{Napier2021,
       author = {{Napier}, K.~J. and {Gerdes}, D.~W. and {Lin}, Hsing Wen and {Hamilton}, S.~J. and {Bernstein}, G.~M. and {Bernardinelli}, P.~H. and {Abbott}, T.~M.~C. and {Aguena}, M. and {Annis}, J. and {Avila}, S. and {Bacon}, D. and {Bertin}, E. and {Brooks}, D. and {Burke}, D.~L. and {Carnero Rosell}, A. and {Carrasco Kind}, M. and {Carretero}, J. and {Costanzi}, M. and {da Costa}, L.~N. and {De Vicente}, J. and {Diehl}, H.~T. and {Doel}, P. and {Everett}, S. and {Ferrero}, I. and {Fosalba}, P. and {Garc{\'\i}a-Bellido}, J. and {Gruen}, D. and {Gruendl}, R.~A. and {Gutierrez}, G. and {Hollowood}, D.~L. and {Honscheid}, K. and {Hoyle}, B. and {James}, D.~J. and {Kent}, S. and {Kuehn}, K. and {Kuropatkin}, N. and {Maia}, M.~A.~G. and {Menanteau}, F. and {Miquel}, R. and {Morgan}, R. and {Palmese}, A. and {Paz-Chinch{\'o}n}, F. and {Plazas}, A.~A. and {Sanchez}, E. and {Scarpine}, V. and {Serrano}, S. and {Sevilla-Noarbe}, I. and {Smith}, M. and {Suchyta}, E. and {Swanson}, M.~E.~C. and {To}, C. and {Walker}, A.~R. and {Wilkinson}, R.~D. and {DES Collaboration}},
        title = "{No Evidence for Orbital Clustering in the Extreme Trans-Neptunian Objects}",
      journal = {\psj},
         year = 2021,
        month = apr,
       volume = {2},
       number = {2},
          eid = {59},
        pages = {59},
          doi = {10.3847/PSJ/abe53e},
archivePrefix = {arXiv},
       eprint = {2102.05601},
 primaryClass = {astro-ph.EP},
       adsurl = {https://ui.adsabs.harvard.edu/abs/2021PSJ.....2...59N}
}

@ARTICLE{SheppardTrujillo2016,
       author = {{Sheppard}, Scott S. and {Trujillo}, Chadwick},
        title = "{New Extreme Trans-Neptunian Objects: Toward a Super-Earth in the Outer Solar System}",
      journal = {\aj},
         year = 2016,
        month = dec,
       volume = {152},
       number = {6},
          eid = {221},
        pages = {221},
          doi = {10.3847/1538-3881/152/6/221},
archivePrefix = {arXiv},
       eprint = {1608.08772},
 primaryClass = {astro-ph.EP},
       adsurl = {https://ui.adsabs.harvard.edu/abs/2016AJ....152..221S}
}

@ARTICLE{DES,
       author = {{The Dark Energy Survey Collaboration}},
        title = "{The Dark Energy Survey}",
      journal = {arXiv e-prints},
         year = 2005,
        month = oct,
          eid = {astro-ph/0510346},
        pages = {astro-ph/0510346},
          doi = {10.48550/arXiv.astro-ph/0510346},
archivePrefix = {arXiv},
       eprint = {astro-ph/0510346},
 primaryClass = {astro-ph},
       adsurl = {https://ui.adsabs.harvard.edu/abs/2005astro.ph.10346T}
}

@ARTICLE{Bannister2016,
       author = {{Bannister}, Michele T. and {Kavelaars}, J.~J. and {Petit}, Jean-Marc and {Gladman}, Brett J. and {Gwyn}, Stephen D.~J. and {Chen}, Ying-Tung and {Volk}, Kathryn and {Alexandersen}, Mike and {Benecchi}, Susan D. and {Delsanti}, Audrey and {Fraser}, Wesley C. and {Granvik}, Mikael and {Grundy}, Will M. and {Guilbert-Lepoutre}, Aur{\'e}lie and {Hestroffer}, Daniel and {Ip}, Wing-Huen and {Jakubik}, Marian and {Jones}, R. Lynne and {Kaib}, Nathan and {Kavelaars}, Catherine F. and {Lacerda}, Pedro and {Lawler}, Samantha and {Lehner}, Matthew J. and {Lin}, Hsing Wen and {Lister}, Tim and {Lykawka}, Patryk Sofia and {Monty}, Stephanie and {Marsset}, Michael and {Murray-Clay}, Ruth and {Noll}, Keith S. and {Parker}, Alex and {Pike}, Rosemary E. and {Rousselot}, Philippe and {Rusk}, David and {Schwamb}, Megan E. and {Shankman}, Cory and {Sicardy}, Bruno and {Vernazza}, Pierre and {Wang}, Shiang-Yu},
        title = "{The Outer Solar System Origins Survey. I. Design and First-quarter Discoveries}",
      journal = {\aj},
         year = 2016,
        month = sep,
       volume = {152},
       number = {3},
          eid = {70},
        pages = {70},
          doi = {10.3847/0004-6256/152/3/70},
archivePrefix = {arXiv},
       eprint = {1511.02895},
 primaryClass = {astro-ph.EP},
       adsurl = {https://ui.adsabs.harvard.edu/abs/2016AJ....152...70B}
}

@ARTICLE{Naess2021,
       author = {{Naess}, Sigurd and {Aiola}, Simone and {Battaglia}, Nick and {Bond}, Richard J. and {Calabrese}, Erminia and {Choi}, Steve K. and {Cothard}, Nicholas F. and {Halpern}, Mark and {Hill}, J. Colin and {Koopman}, Brian J. and {Devlin}, Mark and {McMahon}, Jeff and {Dicker}, Simon and {Duivenvoorden}, Adriaan J. and {Dunkley}, Jo and {Fanfani}, Valentina and {Ferraro}, Simone and {Gallardo}, Patricio A. and {Guan}, Yilun and {Han}, Dongwon and {Hasselfield}, Matthew and {Hincks}, Adam D. and {Huffenberger}, Kevin and {Kosowsky}, Arthur B. and {Louis}, Thibaut and {Macinnis}, Amanda and {Madhavacheril}, Mathew S. and {Nati}, Federico and {Niemack}, Michael D. and {Page}, Lyman and {Salatino}, Maria and {Schaan}, Emmanuel and {Orlowski-Scherer}, John and {Schillaci}, Alessandro and {Schmitt}, Benjamin and {Sehgal}, Neelima and {Sif{\'o}n}, Crist{\'o}bal and {Staggs}, Suzanne and {Engelen}, Alexander Van and {Wollack}, Edward J.},
        title = "{The Atacama Cosmology Telescope: A Search for Planet 9}",
      journal = {\apj},
         year = 2021,
        month = dec,
       volume = {923},
       number = {2},
          eid = {224},
        pages = {224},
          doi = {10.3847/1538-4357/ac2307},
archivePrefix = {arXiv},
       eprint = {2104.10264},
 primaryClass = {astro-ph.EP},
       adsurl = {https://ui.adsabs.harvard.edu/abs/2021ApJ...923..224N}
}

@ARTICLE{Wu2013,
       author = {{Wu}, Yanqin and {Lithwick}, Yoram},
        title = "{Density and Eccentricity of Kepler Planets}",
      journal = {\apj},
         year = 2013,
        month = jul,
       volume = {772},
       number = {1},
          eid = {74},
        pages = {74},
          doi = {10.1088/0004-637X/772/1/74},
archivePrefix = {arXiv},
       eprint = {1210.7810},
 primaryClass = {astro-ph.EP},
       adsurl = {https://ui.adsabs.harvard.edu/abs/2013ApJ...772...74W}
}

@ARTICLE{Miyazaki2018,
       author = {{Miyazaki}, Satoshi and {Komiyama}, Yutaka and {Kawanomoto}, Satoshi and {Doi}, Yoshiyuki and {Furusawa}, Hisanori and {Hamana}, Takashi and {Hayashi}, Yusuke and {Ikeda}, Hiroyuki and {Kamata}, Yukiko and {Karoji}, Hiroshi and {Koike}, Michitaro and {Kurakami}, Tomio and {Miyama}, Shoken and {Morokuma}, Tomoki and {Nakata}, Fumiaki and {Namikawa}, Kazuhito and {Nakaya}, Hidehiko and {Nariai}, Kyoji and {Obuchi}, Yoshiyuki and {Oishi}, Yukie and {Okada}, Norio and {Okura}, Yuki and {Tait}, Philip and {Takata}, Tadafumi and {Tanaka}, Yoko and {Tanaka}, Masayuki and {Terai}, Tsuyoshi and {Tomono}, Daigo and {Uraguchi}, Fumihiro and {Usuda}, Tomonori and {Utsumi}, Yousuke and {Yamada}, Yoshihiko and {Yamanoi}, Hitomi and {Aihara}, Hiroaki and {Fujimori}, Hiroki and {Mineo}, Sogo and {Miyatake}, Hironao and {Oguri}, Masamune and {Uchida}, Tomohisa and {Tanaka}, Manobu M. and {Yasuda}, Naoki and {Takada}, Masahiro and {Murayama}, Hitoshi and {Nishizawa}, Atsushi J. and {Sugiyama}, Naoshi and {Chiba}, Masashi and {Futamase}, Toshifumi and {Wang}, Shiang-Yu and {Chen}, Hsin-Yo and {Ho}, Paul T.~P. and {Liaw}, Eric J.~Y. and {Chiu}, Chi-Fang and {Ho}, Cheng-Lin and {Lai}, Tsang-Chih and {Lee}, Yao-Cheng and {Jeng}, Dun-Zen and {Iwamura}, Satoru and {Armstrong}, Robert and {Bickerton}, Steve and {Bosch}, James and {Gunn}, James E. and {Lupton}, Robert H. and {Loomis}, Craig and {Price}, Paul and {Smith}, Steward and {Strauss}, Michael A. and {Turner}, Edwin L. and {Suzuki}, Hisanori and {Miyazaki}, Yasuhito and {Muramatsu}, Masaharu and {Yamamoto}, Koei and {Endo}, Makoto and {Ezaki}, Yutaka and {Ito}, Noboru and {Kawaguchi}, Noboru and {Sofuku}, Satoshi and {Taniike}, Tomoaki and {Akutsu}, Kotaro and {Dojo}, Naoto and {Kasumi}, Kazuyuki and {Matsuda}, Toru and {Imoto}, Kohei and {Miwa}, Yoshinori and {Suzuki}, Masayuki and {Takeshi}, Kunio and {Yokota}, Hideo},
        title = "{Hyper Suprime-Cam: System design and verification of image quality}",
      journal = {\pasj},
         year = 2018,
        month = jan,
       volume = {70},
          eid = {S1},
        pages = {S1},
          doi = {10.1093/pasj/psx063},
       adsurl = {https://ui.adsabs.harvard.edu/abs/2018PASJ...70S...1M}
}

@article{Taylor_2014,
doi = {10.1088/0004-637X/792/2/135},
url = {https://dx.doi.org/10.1088/0004-637X/792/2/135},
year = {2014},
month = {aug},
publisher = {The American Astronomical Society},
volume = {792},
number = {2},
pages = {135},
author = {Taylor, Matt and Cinabro, David and Dilday, Ben and Galbany, Lluis and Gupta, Ravi R. and Kessler, R. and Marriner, John and Nichol, Robert C. and Richmond, Michael and Schneider, Donald P. and Sollerman, Jesper},
title = {THE CORE COLLAPSE SUPERNOVA RATE FROM THE SDSS-II SUPERNOVA SURVEY},
journal = {The Astrophysical Journal}}

@article{Perley_2022,
doi = {10.3847/1538-4357/ac478e},
url = {https://dx.doi.org/10.3847/1538-4357/ac478e},
year = {2022},
month = {mar},
publisher = {The American Astronomical Society},
volume = {927},
number = {2},
pages = {180},
author = {Perley, Daniel A. and Sollerman, Jesper and Schulze, Steve and Yao, Yuhan and Fremling, Christoffer and Gal-Yam, Avishay and Ho, Anna Y. Q. and Yang, Yi and Kool, Erik C. and Irani, Ido and Yan, Lin and Andreoni, Igor and Baade, Dietrich and Bellm, Eric C. and Brink, Thomas G. and Chen, Ting-Wan and Cikota, Aleksandar and Coughlin, Michael W. and Dahiwale, Aishwarya and Dekany, Richard and Duev, Dmitry A. and Filippenko, Alexei V. and Hoeflich, Peter and Kasliwal, Mansi M. and Kulkarni, S. R. and Lunnan, Ragnhild and Masci, Frank J. and Maund, Justyn R. and Medford, Michael S. and Riddle, Reed and Rosnet, Philippe and Shupe, David L. and Strotjohann, Nora Linn and Tzanidakis, Anastasios and Zheng, WeiKang},
title = {The Type Icn SN 2021csp: Implications for the Origins of the Fastest Supernovae and the Fates of Wolf–Rayet Stars},
journal = {The Astrophysical Journal}}

@article{Frohmaier2019,
    author = {Frohmaier, C and Sullivan, M and Nugent, P E and Smith, M and Dimitriadis, G and Bloom, J S and Cenko, S B and Kasliwal, M M and Kulkarni, S R and Maguire, K and Ofek, E O and Poznanski, D and Quimby, R M},
    title = {The volumetric rate of normal type Ia supernovae in the local Universe discovered by the Palomar Transient Factory},
    journal = {Monthly Notices of the Royal Astronomical Society},
    volume = {486},
    number = {2},
    pages = {2308-2320},
    year = {2019},
    month = {03},
    issn = {0035-8711},
    doi = {10.1093/mnras/stz807},
    url = {https://doi.org/10.1093/mnras/stz807},
    eprint = {https://academic.oup.com/mnras/article-pdf/486/2/2308/28493296/stz807.pdf},
}

@article{Johansson2013,
    author = {Johansson, J. and Amanullah, R. and Goobar, A.},
    title = {Herschel limits on far-infrared emission from circumstellar dust around three nearby Type Ia supernovae},
    journal = {Monthly Notices of the Royal Astronomical Society: Letters},
    volume = {431},
    number = {1},
    pages = {L43-L47},
    year = {2013},
    month = {02},
    issn = {1745-3925},
    doi = {10.1093/mnrasl/slt005},
    url = {https://doi.org/10.1093/mnrasl/slt005},
    eprint = {https://academic.oup.com/mnrasl/article-pdf/431/1/L43/54661361/mnrasl\_431\_1\_l43.pdf},
}

@article{Masterson_2024,
doi = {10.3847/1538-4357/ad18bb},
url = {https://dx.doi.org/10.3847/1538-4357/ad18bb},
year = {2024},
month = {jan},
publisher = {The American Astronomical Society},
volume = {961},
number = {2},
pages = {211},
author = {Masterson, Megan and De, Kishalay and Panagiotou, Christos and Kara, Erin and Arcavi, Iair and Eilers, Anna-Christina and Frostig, Danielle and Gezari, Suvi and Grotova, Iuliia and Liu, Zhu and Malyali, Adam and Meisner, Aaron M. and Merloni, Andrea and Newsome, Megan and Rau, Arne and Simcoe, Robert A. and van Velzen, Sjoert},
title = {A New Population of Mid-infrared-selected Tidal Disruption Events: Implications for Tidal Disruption Event Rates and Host Galaxy Properties},
journal = {The Astrophysical Journal}}

@article{Kawash_2021,
doi = {10.3847/1538-4357/ac1f1a},
url = {https://dx.doi.org/10.3847/1538-4357/ac1f1a},
year = {2021},
month = {nov},
publisher = {The American Astronomical Society},
volume = {922},
number = {1},
pages = {25},
author = {Kawash, A. and Chomiuk, L. and Rodriguez, J. A. and Strader, J. and Sokolovsky, K. V. and Aydi, E. and Kochanek, C. S. and Stanek, K. Z. and Mukai, K. and De, K. and Shappee, B. and Holoien, T. W.-S. and Prieto, J. L. and Thompson, T. A.},
title = {Galactic Extinction: How Many Novae Does It Hide and How Does It Affect the Galactic Nova Rate?},
journal = {The Astrophysical Journal}}

@article{Planck2016,
   title={Planckintermediate results: XLVII.Planckconstraints on reionization history},
   volume={596},
   ISSN={1432-0746},
   url={http://dx.doi.org/10.1051/0004-6361/201628897},
   DOI={10.1051/0004-6361/201628897},
   publisher={EDP Sciences},
   author={Adam, R. and Aghanim, N. and Ashdown, M. and Aumont, J. and Baccigalupi, C. and Ballardini, M. and Banday, A. J. and Barreiro, R. B. and Bartolo, N. and Basak, S. and Battye, R. and Benabed, K. and Bernard, J.-P. and Bersanelli, M. and Bielewicz, P. and Bock, J. J. and Bonaldi, A. and Bonavera, L. and Bond, J. R. and Borrill, J. and Bouchet, F. R. and Boulanger, F. and Bucher, M. and Burigana, C. and Calabrese, E. and Cardoso, J.-F. and Carron, J. and Chiang, H. C. and Colombo, L. P. L. and Combet, C. and Comis, B. and Couchot, F. and Coulais, A. and Crill, B. P. and Curto, A. and Cuttaia, F. and Davis, R. J. and de Bernardis, P. and de Rosa, A. and de Zotti, G. and Delabrouille, J. and Di Valentino, E. and Dickinson, C. and Diego, J. M. and Doré, O. and Douspis, M. and Ducout, A. and Dupac, X. and Elsner, F. and Enßlin, T. A. and Eriksen, H. K. and Falgarone, E. and Fantaye, Y. and Finelli, F. and Forastieri, F. and Frailis, M. and Fraisse, A. A. and Franceschi, E. and Frolov, A. and Galeotta, S. and Galli, S. and Ganga, K. and Génova-Santos, R. T. and Gerbino, M. and Ghosh, T. and González-Nuevo, J. and Górski, K. M. and Gruppuso, A. and Gudmundsson, J. E. and Hansen, F. K. and Helou, G. and Henrot-Versillé, S. and Herranz, D. and Hivon, E. and Huang, Z. and Ilić, S. and Jaffe, A. H. and Jones, W. C. and Keihänen, E. and Keskitalo, R. and Kisner, T. S. and Knox, L. and Krachmalnicoff, N. and Kunz, M. and Kurki-Suonio, H. and Lagache, G. and Lähteenmäki, A. and Lamarre, J.-M. and Langer, M. and Lasenby, A. and Lattanzi, M. and Lawrence, C. R. and Le Jeune, M. and Levrier, F. and Lewis, A. and Liguori, M. and Lilje, P. B. and López-Caniego, M. and Ma, Y.-Z. and Macías-Pérez, J. F. and Maggio, G. and Mangilli, A. and Maris, M. and Martin, P. G. and Martínez-González, E. and Matarrese, S. and Mauri, N. and McEwen, J. D. and Meinhold, P. R. and Melchiorri, A. and Mennella, A. and Migliaccio, M. and Miville-Deschênes, M.-A. and Molinari, D. and Moneti, A. and Montier, L. and Morgante, G. and Moss, A. and Naselsky, P. and Natoli, P. and Oxborrow, C. A. and Pagano, L. and Paoletti, D. and Partridge, B. and Patanchon, G. and Patrizii, L. and Perdereau, O. and Perotto, L. and Pettorino, V. and Piacentini, F. and Plaszczynski, S. and Polastri, L. and Polenta, G. and Puget, J.-L. and Rachen, J. P. and Racine, B. and Reinecke, M. and Remazeilles, M. and Renzi, A. and Rocha, G. and Rossetti, M. and Roudier, G. and Rubiño-Martín, J. A. and Ruiz-Granados, B. and Salvati, L. and Sandri, M. and Savelainen, M. and Scott, D. and Sirri, G. and Sunyaev, R. and Suur-Uski, A.-S. and Tauber, J. A. and Tenti, M. and Toffolatti, L. and Tomasi, M. and Tristram, M. and Trombetti, T. and Valiviita, J. and Van Tent, F. and Vielva, P. and Villa, F. and Vittorio, N. and Wandelt, B. D. and Wehus, I. K. and White, M. and Zacchei, A. and Zonca, A.},
   year={2016},
   month=dec, pages={A108} }

\appendix

\end{document}